\documentclass{aa}  

\usepackage{graphicx}
\usepackage{subfigure}
\usepackage{grffile}
\usepackage{txfonts}
\usepackage{hyperref}
\def\<#1>{\langle#1\rangle}

\begin{document} 
\graphicspath{{figures/}}

   \title{Subarcsecond international LOFAR radio images of \\ Arp\,220 at 150\,MHz}

	\subtitle{A kpc-scale star forming disk surrounding nuclei with shocked outflows}

      \author{E. Varenius \inst{\ref{inst:chalmers}}
          \and
          J.~E. Conway \inst{\ref{inst:chalmers}}
          \and
          I. Martí-Vidal \inst{\ref{inst:chalmers}}
          \and
          S. Aalto \inst{\ref{inst:chalmers}}
          \and
          L. Barcos-Mu\~noz \inst{\ref{inst:loreto}}
          \and
          S. K{\"o}nig \inst{\ref{inst:chalmers}}
          \and 
          M.~A. P\'erez-Torres \inst{\ref{inst:iaa},\ref{inst:unizar}}
          \and \\
          A.~T. Deller \inst{\ref{inst:ASTRON}}
		  \and
		  J. Moldón \inst{\ref{inst:UK}}
          \and
          J. S. Gallagher \inst{\ref{inst:jay}}
          \and
          T.~M. Yoast-Hull \inst{\ref{inst:tova1},\ref{inst:tova2}}
          \and
          C. Horellou \inst{\ref{inst:chalmers}}
          \and
          L.~K. Morabito \inst{\ref{inst:leiden}}
          \and
          A. Alberdi \inst{\ref{inst:iaa}}
          \and \\
          N. Jackson \inst{\ref{inst:UK}}
          \and
          R. Beswick \inst{\ref{inst:UK}}
          \and
          T.~D. Carozzi \inst{\ref{inst:chalmers}}
          \and
          O. Wucknitz \inst{\ref{inst:mpifr}}
          \and
          N. Ramírez-Olivencia \inst{\ref{inst:iaa}}
          }

   \institute{
              Department of Earth and Space Sciences,
              Chalmers University of Technology, 
              Onsala Space Observatory,
              439 92 Onsala, 
              Sweden
              \email{varenius@chalmers.se}
              \label{inst:chalmers}
              \and
			  Department of Astronomy, University of Virginia, 
              530 McCormick Road, Charlottesville, VA 22904, USA
              \label{inst:loreto}
              \and
              The Netherlands Institute for Radio Astronomy (ASTRON), PO Box 2, 7990 AA Dwingeloo, The Netherlands
              \label{inst:ASTRON}
              \and
              Jodrell Bank Centre for Astrophysics,
              Alan Turing Building,
              School of Physics and Astronomy,
              The University of Manchester,
              Manchester M13 9PL,
              UK
              \label{inst:UK}
              \and
			  Instituto de Astrof\'isica de Andaluc\'ia (IAA, CSIC), Glorieta de las Astronom\'ia, s/n, E-18008 Granada, Spain.
              \label{inst:iaa}
              \and
              Departamento de F\'isica Teorica, Facultad de Ciencias, Universidad de Zaragoza, Spain.
              \label{inst:unizar}
              \and
              Department of Astronomy, University of Wisconsin-Madison, WI 53706, USA
              \label{inst:jay}
              \and
              Department of Physics, University of Wisconsin-Madison, WI 53706, USA
              \label{inst:tova1}
              \and
              Center for Magnetic Self-Organization in Laboratory and Astrophysical Plasmas, University of Wisconsin-Madison, WI 53706, USA
              \label{inst:tova2}
              \and
              Leiden Observatory, Leiden University, P.O. Box 9513, NL-2300 RA Leiden, the Netherlands
              \label{inst:leiden}
              \and
              Max-Planck-Institut f\"ur Radioastronomie, Auf dem H\"ugel 69, D-53121 Bonn, Germany
              \label{inst:mpifr}
		  }

   \date{Received April 13, 2016; accepted July 7, 2016}

 
  \abstract
  { Arp\,220 is the prototypical ultra luminous infrared galaxy (ULIRG). Despite 
extensive studies, the structure at MHz-frequencies has remained unknown
because of limits in spatial resolution.
   }
   {
	   This work aims to constrain the flux and shape of radio emission from Arp\,220
at MHz frequencies.
   }
   {
	   We analyse new observations with the International Low Frequency Array
	   (LOFAR) telescope, and archival data from the Multi-Element Radio Linked
	   Interferometer Network (MERLIN) and the Karl G. Jansky Very Large Array
   (VLA). We model the spatially resolved radio spectrum of Arp\,220 from
   150\,MHz to 33\,GHz. 
   }
   {
	   We present an image of Arp\,220 at 150\,MHz with resolution
$0\farcs65\times0\farcs35$, sensitivity 0.15\,mJy~beam$^{-1}$, and integrated
flux density $394\pm59$\,mJy.  More than 80\% of the detected flux comes from
extended ($6''\approx$2.2\,kpc) steep spectrum ($\alpha=-0.7$) emission, likely
from star formation in the molecular disk surrounding the two nuclei. We find
elongated features extending $0.3''$ (110\,pc) and $0.9''$ (330\,pc) from the
eastern and western nucleus respectively, which we interpret as evidence for outflows. The
extent of radio emission requires acceleration of cosmic rays far outside the
nuclei.  We find that a simple three component model can explain most of the
observed radio spectrum of the galaxy. When accounting for absorption at
1.4\,GHz, Arp\,220 follows the FIR/radio correlation with $q=2.36$, and we
estimate a star formation rate of 220\,M$_\odot\text{\,yr}^{-1}$. We derive
thermal fractions at 1\,GHz of less than 1\% for the nuclei, which indicates
that a major part of the UV-photons are absorbed by dust.
   }
   {
	International LOFAR observations shows great promise to detect steep
	spectrum outflows and probe regions of thermal absorption. However, in
	LIRGs the emission detected at 150\,MHz does not necessarily come from the
	main regions of star formation.  This implies that high spatial resolution
	is crucial for accurate estimates of star formation rates for such galaxies
	at 150\,MHz. 
   }


   \maketitle
%

\section{Introduction}
Arp\,220 is the closest (77\,Mpc) ULIRG and has been extensively
studied across the electromagnetic spectrum. It is a late-stage
merger, which explains the peculiar morphology noted by \cite{arp1966}.
The centre is heavily obscured in the optical, but radio observations
reveal two bright sources about 1$''$ (370\,pc) apart, thought to be the
nuclei of two merging galaxies \citep{norris1988}.  The two nuclei resemble
rotating exponential disks with ongoing star formation
\citep{sakamoto1999,sakamoto2008,scoville2015,barcosmunoz2015}, giving rise to
dozens of supernovae and supernova remnants detected using very long baseline
interferometry (VLBI) at cm
wavelengths \citep{smith1998,lonsdale2006,parra2007,batejat2011}.  While most of
the dust and GHz radio continuum comes from the two nuclei
\citep{sakamoto2008,barcosmunoz2015}, CO observations reveal extended dense
molecular gas surrounding the nuclei in a kpc-scale ring or disk
\citep{scoville1997,downes1998,sakamoto2008,koenig2012}.  Optical and X-ray
observations provide evidence for a bi-conical outflow, or superwind, carrying
energy and matter out to several kpc from the centre of the galaxy
\citep{heckman1990,arribas2001,mcdowell2003}. 

While there is plenty of evidence for intense star formation in Arp\,220, the
presence of one or more active galactic nuclei (AGN) has not been established.
\cite{yun2001} finds Arp\,220 fainter at 1.4\,GHz than
expected from the FIR/radio correlation for star forming galaxies.  This may be because
part of the IR comes from a radio-weak AGN, which could also explain the
lack of a clear AGN-candidate amongst the compact objects detected with
cm-VLBI. Unfortunately the extreme column densities of $10^{25}$cm$^{-2}$
\citep{wilson2014} towards the nuclei greatly hinders X-ray observations which
could offer direct evidence of AGN activity. 

The lack of radio emission could also be explained if the synchrotron emission
from the nuclei is significantly reduced by thermal (free-free) absorption at
GHz frequencies. Since this effect is most prominent at lower frequencies (see
e.g. \citealt{condon1992}), observations at MHz frequencies may constrain the
properties and structure of the absorbing medium.

Regardless of the presence of an AGN in Arp\,220, the mechanical energy from
the central starburst would likely contribute significantly to the large-scale
superwind. This could manifest itself as outflows from the nuclei and  
the surrounding molecular disk. Indeed, evidence of outflows from the nuclei 
with speeds of a few hundred km~s$^{-1}$ has been reported
\citep{sakamoto2009,tunnard2015}.
These outflows could carry synchrotron-emitting cosmic rays (CRs) out from the
nuclei where they could potentially be observed at radio frequencies.
Star formation in the molecular disk would also give rise to radio emission
on kpc-scales, i.e. far outside the nuclei.
The intrinsic steep spectrum of unabsorbed synchrotron emission
makes a detection of weak extended emission challenging at GHz frequencies, and
indeed no detection of kpc-scale radio emission has been reported.

Although Arp\,220 has been observed at MHz frequencies before
\citep{sopp1991,waldram1996,douglas1996}, none of those studies resolved the galaxy.
Subarcsecond resolution is crucial to understand the relative contributions
from the emitting and absorbing structures in the centre of Arp\,220.

In this paper, we, for the first time, resolve Arp\,220 at 150\,MHz using the
international LOFAR telescope. To complement these observations, we 
combine archival data from the VLA and MERLIN arrays to obtain both high
resolution and sensitivity to extended structure at 1.4\,GHz. In Sect.
\ref{sect:obs} we describe the data used in this paper. In Sect.
\ref{sect:results} we present our results from the observations, in Sect.
\ref{sect:modeling} we describe our modelling, and in Sect.
\ref{sect:discussion}  we discuss the results and modelling in the context of
previous studies.  Finally, in Sect. \ref{sect:summary}, we summarise our
conclusions.  

Throughout this paper we assume a distance to Arp 220 of 77\,Mpc
($H_0$=70~km~s$^{-1}$\,Mpc$^{-1}$), i.e.  $0\farcs1$=37\,pc, and all spectral
indices, $\alpha$, are given according to the power law $S_\nu\propto
\nu^\alpha$.  

\section{Observations and data reduction}
\label{sect:obs}
In this section we describe the data used in this paper. The focus of this
paper is new data obtained at 150\,MHz using the International LOFAR telescope.
These data are briefly described in Sect. \ref{sect:obslofar}, and the
interested reader can find more details in Appendix \ref{app:lofarcal}. To
complement our LOFAR observations, we also present a new image obtained at
1.4\,GHz by combining archival data from the VLA and MERLIN arrays, as described in
Sect.  \ref{sect:vla+merlin}. Finally, in Sect. \ref{sect:restobs}, we briefly
describe five additional images included for comparison: radio
continuum at 6\,GHz and 33\,GHz from the VLA, CO(1-0) and CO(2-1) from the
IRAM Plateau de Bure interferometer, and an optical image from the {\it Hubble} Space
Telescope (HST). 

\subsection{LOFAR at 150\,MHz}
\label{sect:obslofar}
We present new data from the International LOFAR telescope, project LC2\_042
(P.I.: E. Varenius). These data were taken between 18:30 and 02:15 UT (with six
hours on Arp\,220) on June 4th 2014, including 46 LOFAR high band array (HBA)
stations: 24 core stations (CS) in \emph{joined} mode (where the two 24-tile
``ears'' of the station are added to form a single station), 14 remote
stations (RS), and eight international stations (IS).  The total available
bandwidth was split equally between two simultaneous beams of width 48\,MHz (240 sub-bands),
centred on 150\,MHz; one on Arp\,220 and one on the bright and compact
calibrator J1513+2388 separated from Arp\,220 by $4.9^\circ$.  Every 20 minutes
the observations switched, for 2 minutes to a single beam on the absolute flux
density calibrator 3C295 separated from Arp\,220 by 32$^\circ$.  
Although in theory a single scan of 3C295 is sufficient to determine the flux
scale, we included scans every 20 minutes to a) track residual beam effects
over time and b), track phase variations between CS to enable coherent addition
of CS to increase sensitivity in case J1513+2388 was weaker than expected.
While no addition of CS was necessary for calibration, we found significant
beam effects which limited the time range used for setting the flux scale and
spectral index. We carefully selected a time range where these beam effects
cause amplitude uncertainties of at most 10\,\%.
Based on the measured gain variations as function of elevation angle, and the
uncertainties on the flux scale as given by \cite{scaife2012}, we estimate
the flux density to be accurate to within 15\,\%.

Residual delays and rates, caused by the e.g. the ionosphere and imperfect
station clocks, are significant on international baselines and have to be
removed before further calibration and imaging. It is challenging to find a nearby (few degrees
to track the ionosphere) source which is bright enough on subarcsecond scales
(for international baselines) to derive phase corrections for each channel (or
sub-band) \citep{moldon2014}. Instead we determined
residual delays and rates towards J1513+2388 by combining sub-bands together
assuming a piece-wise linear approximation for eight blocks of 5.9\,MHz each,
a strategy similar to the one used by \cite{varenius2015} for M82.

Residual phase errors were
corrected by hybrid imaging of Arp\,220. This method solves for relative phase
errors, but does not give absolute phase information.  The absolute position of
Arp\,220 was therefore determined by matching the positions of two compact sources
detected 5' from Arp\,220 both at 150\,MHz and 1.4\,GHz. We conservatively
estimate our positional uncertainty to be at most $\pm68$\,mas in R.A. and
$\pm83$\,mas in Dec.
The data were calibrated in AIPS 31DEC15 \citep{greisen} using ParselTongue 2.0
\citep{Kettenis}.  Using the multi-scale CLEAN algorithm, as implemented in
CASA 4.5.2 \citep{CASA}, we obtained a continuum image of Arp\,220 at 150\,MHz
with resolution of $0\farcs65\times0\farcs35$ and RMS noise
0.15\,mJy~beam$^{-1}$ (see Fig.~\ref{fig:LOFAR}). 
Further details about calibration and
imaging of these data can be found in Appendix \ref{app:lofarcal}.

\begin{figure*}[htbp]
\centering
\subfigure[LOFAR 150\,MHz continuum]{
        \includegraphics[width=0.48\textwidth]{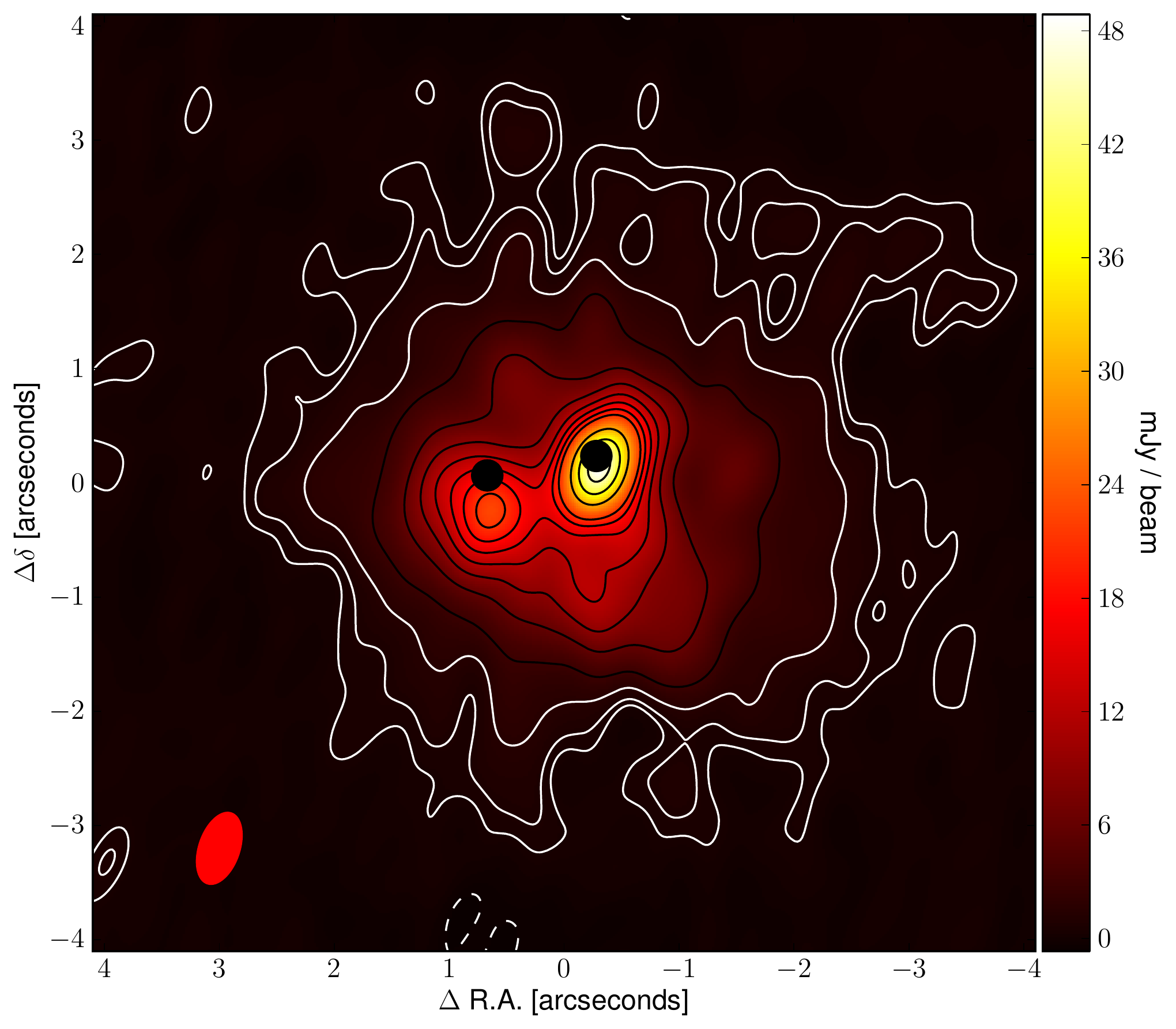}
        \label{fig:LOFAR}
}
\subfigure[Sketch of structure discussed in this paper]{
        \includegraphics[width=0.48\textwidth]{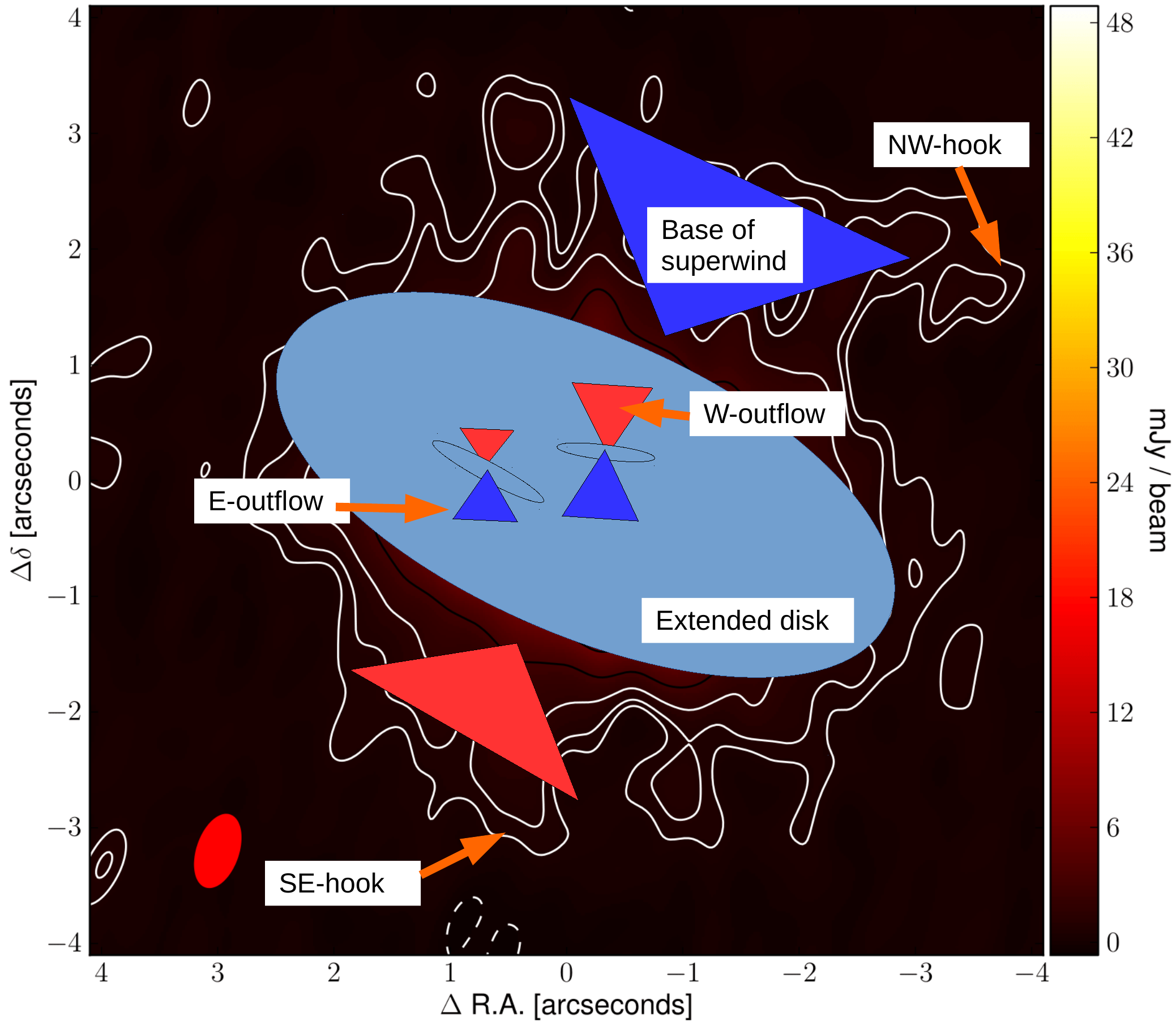}
        \label{fig:sketch}
}
\caption{Left panel \subref{fig:LOFAR}: The continuum emission at 150\,MHz as
	detected with LOFAR, with contours at
	[$-3$,3,5,10,20,40,60,80,100,120,140,200,250,300]$\times$0.15\,mJy~beam$^{-1}$ plotted
	in white in darker areas and black in brighter areas.
	The CLEAN restoring beam of $0\farcs65\times0\farcs35$, position angle 161$^\circ$,
	is plotted as a solid red ellipse in the lower left. Positions are relative to R.A.
	$15^{\rm h}34^{\rm m}57^{\rm s}.2424$, Dec.  $23^\circ30'11\farcs275$. The
black dots in the centre mark the 33\,GHz continuum positions of the two nuclei
\citep{barcosmunoz2015} with radius 133\,mas illustrating the total maximum astrometric
uncertainty between the 150\,MHz (83\,mas) and 33\,GHz (50\,mas).  Right panel
\subref{fig:sketch}: A sketch of the structure discussed in this paper.  The
labels mark the extended kpc-scale disk, the outflows from the east and west
nuclei (small ellipses), the large scale outflow and the north-west and
south-east ``hooks''. The blue triangles represent the closer sides of outflows
and the red triangles the far sides.  \label{fig:LOFAR+sketch}
}
\end{figure*}
   

   \begin{table}
      \caption[]{List of positions of objects referred to in this work.}
         \label{tab:targetlist}
         \begin{tabular}{ l l l}
            Source      &  R. A. [J2000] & Dec. [J2000]\\
            \hline
            3C295 \tablefootmark{a}       & $14^{\rm h}11^{\rm m}20^{\rm s}.50$ & $52^\circ12'10\farcs0$\\
			Arp\,220 \tablefootmark{b}    & $15^{\rm h}34^{\rm m}57^{\rm s}.25$ & $23^\circ30'11\farcs3$ \\
			J1513+2388 \tablefootmark{c}  & $15^{\rm h}13^{\rm m}40^{\rm s}.1857$& $23^\circ38'35\farcs201$\\
			1551+239  \tablefootmark{c} & $15^{\rm h}53^{\rm m}43^{\rm s}.5913$ & $23^\circ48'25\farcs458$\\
            \noalign{\smallskip}
            \hline
         \end{tabular}
         \tablefoot{
             \tablefoottext{a}{From NED, http://ned.ipac.caltech.edu.}
			 \tablefoottext{b}{LOFAR correlation position.}
			 \tablefoottext{c}{Archival VLA/MERLIN position was updated using the \emph{rfc\_2015d} catalogue available
			 via http://astrogeo.org/calib/search/html.}
}
   \end{table}

\subsection{VLA and MERLIN at 1.4\,GHz}
\label{sect:vla+merlin}
To complement our LOFAR data we re-calibrated and combined archival
data from MERLIN and VLA at 1.4\,GHz.  In this section we describe the 
calibration and imaging of these data.

\subsubsection{VLA at 1.4\,GHz}
\label{sect:vladata}
We used archival VLA-data observed on March 25th 1998 (project AA216) in
A-configuration with 5.5 hours on Arp\,220. The 12.5\,MHz bandwidth was split
into two spectral windows of 31 channels each, centred on 1374\,MHz and
1424\,MHz. The data were calibrated in a standard manner using ParselTongue 2.0
\citep{Kettenis} and AIPS 31DEC15 \citep{greisen}.  Phase corrections were
first derived using J1513+2338, assuming the position in the \emph{rfc\_2015d}
catalogue (see Table \ref{tab:targetlist}), and refined using self-calibration
of Arp\,220.  The flux scale was set by 3C286, assuming 15.3\,Jy and 15.0\,Jy
in the two respective spectral windows (as calculated by \verb!SETJY! in AIPS).
We adopt an absolute flux density uncertainty of 10\% for these data.

\subsubsection{MERLIN at 1.4\,GHz}
We used archival MERLIN data taken on January 30th 1996 with 8.3 hours on
Arp\,220, previously published by \cite{mundell2001}.  We choose
to re-calibrate the data to be able to combine them in Fourier space with the VLA
data described in Sect. \ref{sect:vladata}.  The MERLIN data cover 7.8\,MHz
bandwidth split in 63 channels, centred on 1.42\,GHz where Arp\,220
has deep HI absorption.  The time resolution was 16 seconds per sample.  The
data were calibrated in a standard manner in AIPS 31DEC15 \citep{greisen} using
ParselTongue 2.0 \citep{Kettenis}.  Phase corrections were first derived using
1551+239, assuming the position in the \emph{rfc\_2015d} catalogue (see Table
\ref{tab:targetlist}), and refined using self-calibration of Arp\,220. The flux
scale was set by 3C286 using a model supplied by the eMERLIN staff to account
for resolution effects where the total flux density was set to 14.98\,Jy.  The
visibility weights were re-calculated to reflect the RMS scatter within 10
minute intervals.  We adopt an absolute flux density uncertainty of 10\% for
MERLIN.  To avoid effects of HI absorption known to be significant towards the
nuclei, we, after inspection of which channels showed strong HI absorption,
excluded the ten middle channels (1.25\,MHz) of the MERLIN data when imaging
the combined VLA+MERLIN data as described below.

\subsubsection{Combining VLA and MERLIN}
To achieve, at 1.4\,GHz, both sensitivity to extended emission as well as
subarcsecond resolution, we combined the calibrated UV-data from VLA and MERLIN
into one measurement set using the task \verb!concat! in CASA. Using the CLEAN
algorithm as implemented in CASA 4.5.2, with robustness parameter $-0.25$
\citep{briggs} we obtained an image with resolution $0\farcs46\times0\farcs33$  and
RMS noise $\sigma=60\,\mu$Jy~beam$^{-1}$. 

\subsection{Additional images included for interpretation}
\label{sect:restobs}
We include five additional images obtained by previous studies (i.e., we did not
remake the images ourselves).
\begin{figure*}[htbp]
\centering
\subfigure[150\,MHz (black) and 1.4\,GHz (red)]{
        \includegraphics[width=0.48\textwidth]{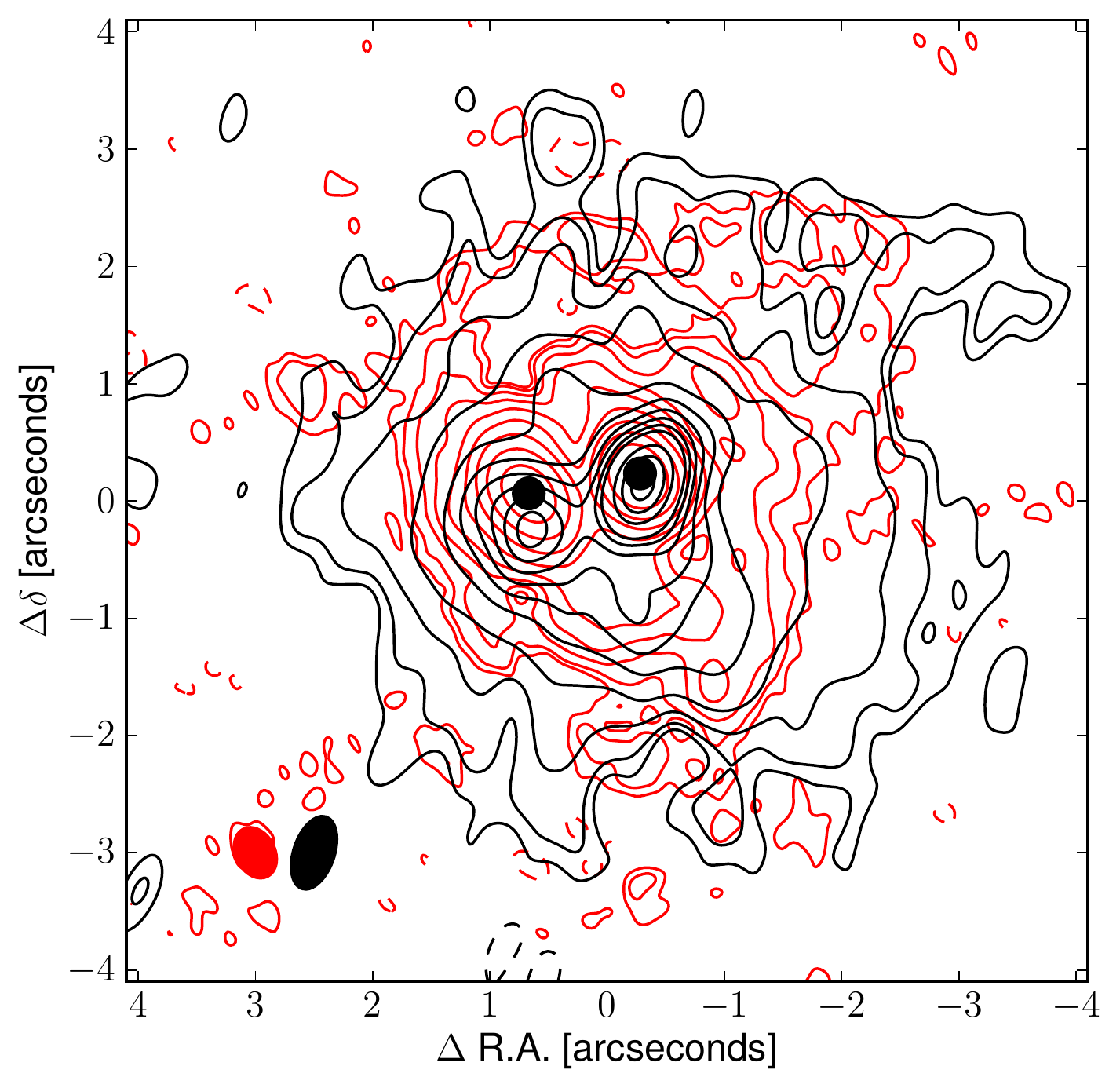}
        \label{fig:lofarmerlin}
}
\subfigure[150\,MHz (black) and 6.0\,GHz (red)]{
        \includegraphics[width=0.48\textwidth]{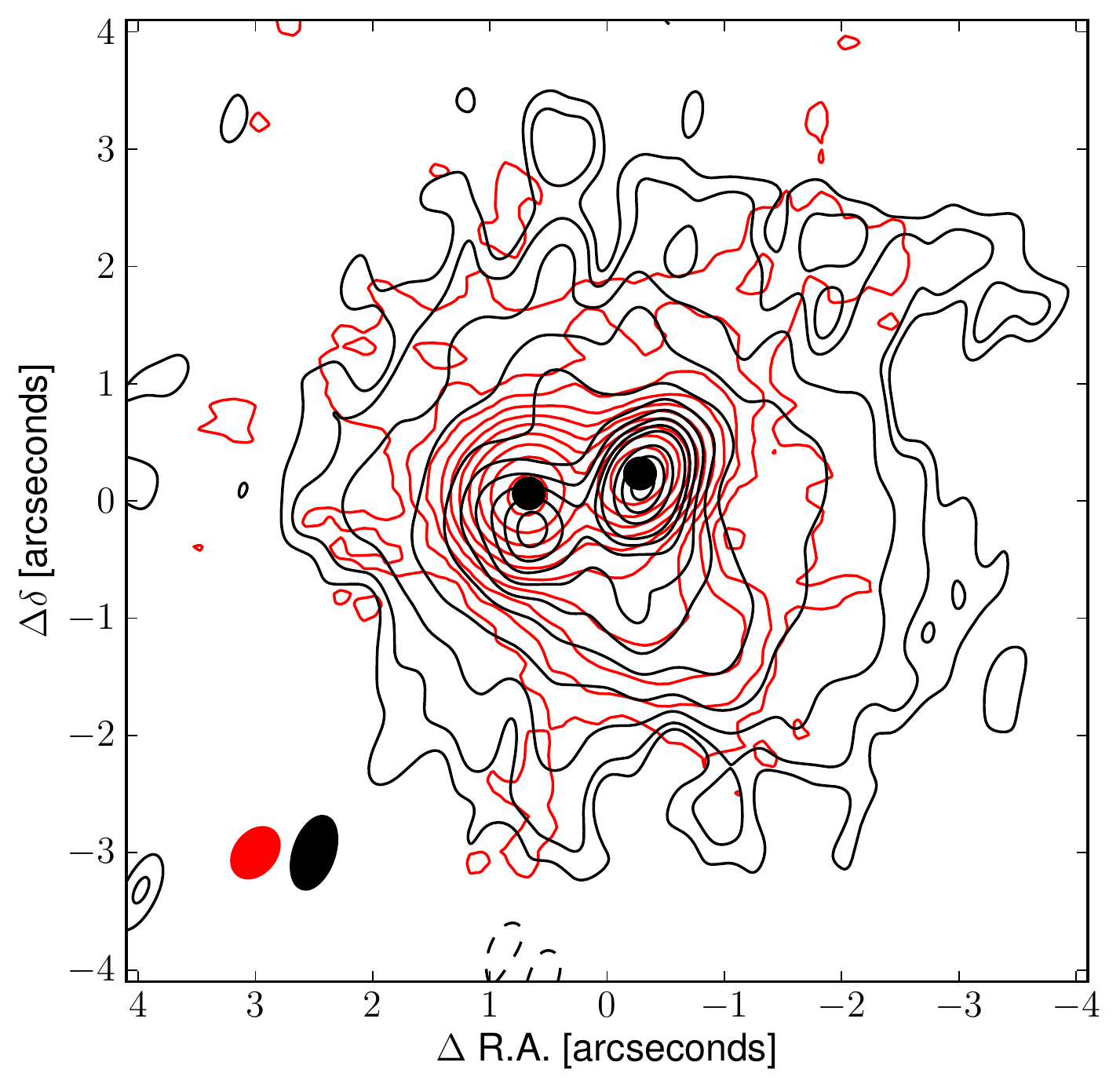}
        \label{fig:lofarvla}
}
\subfigure[150\,MHz (black) and CO 1-0 (red)]{
        \includegraphics[width=0.48\textwidth]{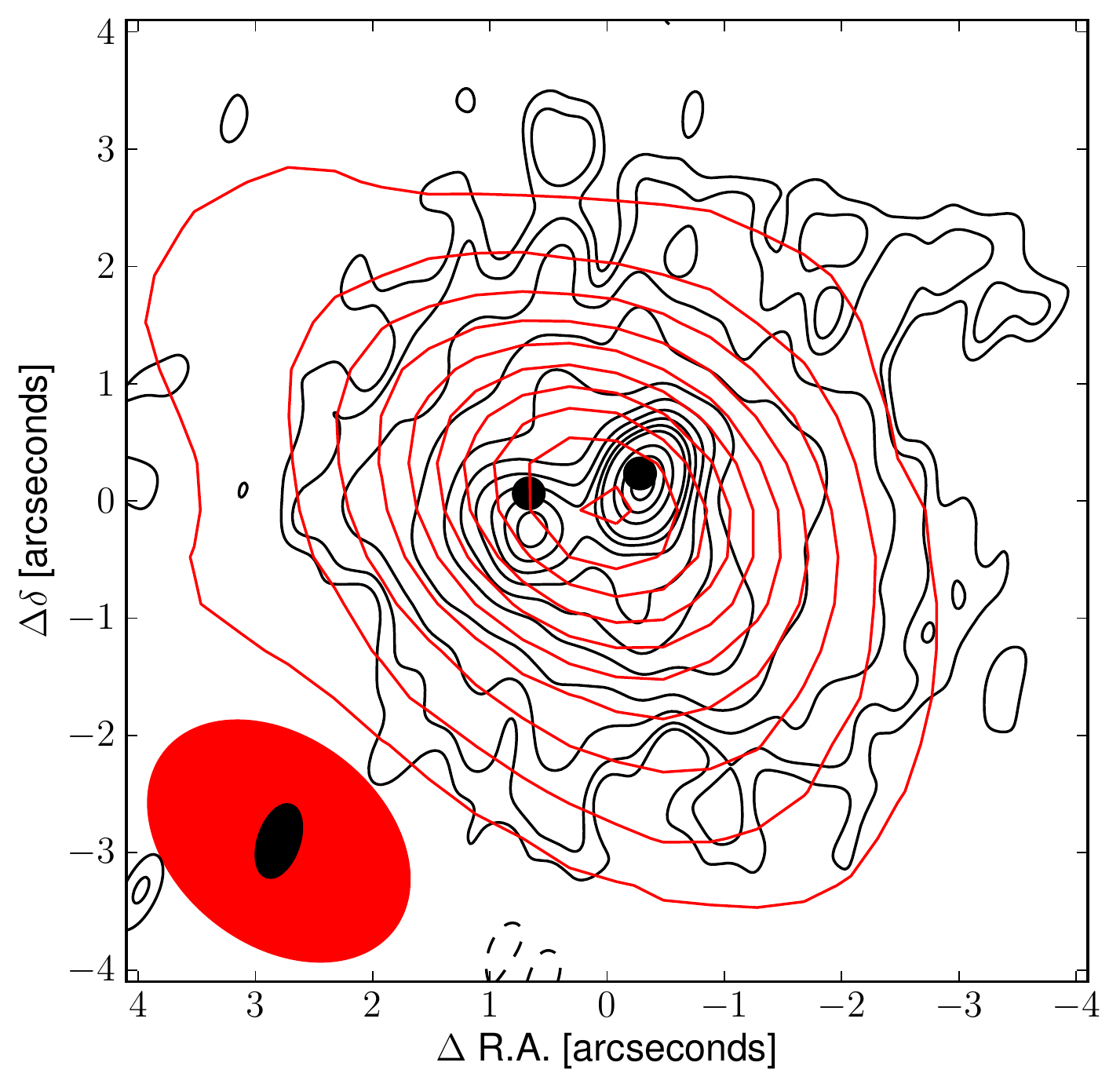}
        \label{fig:co10}
}
\subfigure[150\,MHz (black) and CO 2-1 (red)]{
        \includegraphics[width=0.48\textwidth]{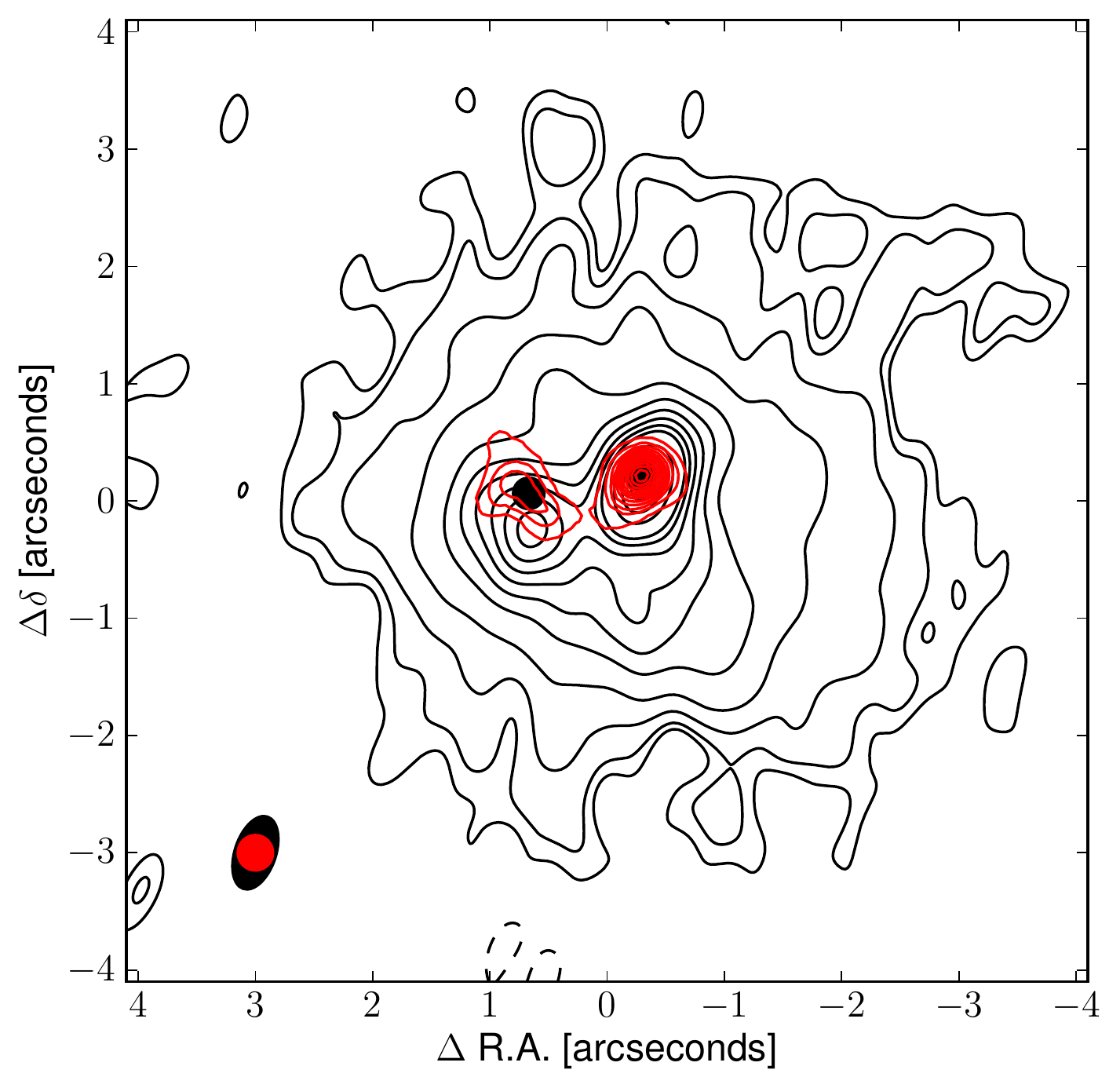}
        \label{fig:co21}
}
\caption{Comparison of the emission detected at 150\,MHz with emission at four
	other wavelengths. Black contours represent the 150\,MHz emission using the
	same contours as in Fig.~\ref{fig:LOFAR}. Panel \subref{fig:lofarmerlin} shows
	the combined VLA+MERLIN 1.4\,GHz continuum as contours at [$-3$,3,
	5,10,20,40,80,160,320,640]$\times$60\,$\mu$Jy~beam$^{-1}$.  Panel
	\subref{fig:lofarvla} shows the VLA 6\,GHz continuum as contours at
	[$-5$,5,10,20,40,80,160,320,640,1280,2560] $\times$14\,$\mu$Jy~beam$^{-1}$.
	Panel \subref{fig:co10} shows CO(1-0) as contours at
	[10,50,90,130,170,210,250,290,330,370]$\times$13.906\,K~kms$^{-1}$.  Panel
	\subref{fig:co21} shows CO(2-1) as contours at [2, 4, 6, 8, 10, 12, 14, 16,
	18, 20, 22, 24]$\times$1184.69\,K~kms$^{-1}$.  The respective CLEAN
	restoring beams are plotted in the lower left of each panel in red:
	\subref{fig:lofarmerlin} $0\farcs65\times0\farcs35$, \subref{fig:lofarvla}
	$0\farcs48\times0\farcs35$, \subref{fig:co10} $2\farcs46\times1\farcs76$,
	and \subref{fig:co21} $0\farcs30\times0\farcs30$ .  In all panels the LOFAR
	beam of $0\farcs65\times0\farcs35$ is plotted in black.  The black dots
    in the centre mark the 33\,GHz positions as in Fig.~\ref{fig:LOFAR}.
\label{fig:overlays}
}
\end{figure*}

For comparison of primarily the extended emission detected at 150\,MHz, we
include a radio continuum image at 6\,GHz published by
\cite{barcosmunoz2015}. 
We note that although the 6\,GHz
image is very sensitive with RMS noise $\sigma=$14\,$\mu$Jy~beam$^{-1}$ (where
the beam is $0\farcs48\times0\farcs35$), \cite{barcosmunoz2015} do not plot contours
lower than 12.5$\sigma$ in their Fig.~1 since their analysis focuses on the
nuclei. We are however interested also in the weaker extended emission and therefore
show also fainter contours in our Fig.~\ref{fig:lofarvla}.  To compute
pixel-wise spectral index maps, we used the task \verb! imregrid! in CASA with
linear interpolation to decrease the pixel size of the 6\,GHz image from
$0\farcs1$ to the $0\farcs02$ used elsewhere in this paper. 

For modelling of the spectra of the nuclei, we also use the image obtained at
33\,GHz by \cite{barcosmunoz2015}, their Fig.~1c, with resolution
$0\farcs081\times0\farcs063$ and RMS noise 23\,$\mu$Jy~beam$^{-1}$.
The positions of the nuclear disks, as fitted by \cite{barcosmunoz2015}
using this 33\,GHz image, are plotted as black circles in multiple figures in
this paper to guide the eye when comparing the structure at different
frequencies.

For comparison to the structure of molecular gas, we include moment zero images
of CO(1-0) and CO(2-1) from the IRAM Plateau de Bure interferometer, previously
described by \cite{koenig2012}.  The CO(1-0) image is shown
as red contours in Fig.~\ref{fig:co10}, with resolution $2\farcs46\times1\farcs76$
and RMS noise 13.906\,K~kms$^{-1}$.The CO(2-1) image is shown as red contours in
Fig.~\ref{fig:co21}, with resolution $0\farcs3\times0\farcs3$ and RMS noise
1184.69\,K~kms$^{-1}$.

Finally, for comparison with optical data, we include an
archival HST image taken on 2006-01-06 with ACS WFC F814W \footnote{Obtained in
March 2016 from https://archive.stsci.edu/hst/} as shown in Fig.
\ref{fig:HST}. 

\section{Results}
\label{sect:results}

   \begin{table}
      \caption[]{Integrated flux densities for Arp\,220 measured in this work.}
         \label{tab:imageparams}
         \begin{tabular}{ l l r l r}
			 Source &  Obs. freq& Resolution & RMS & Int. flux\\
						  &  [GHz] & [arcsec] & [$\mu$Jy/b]& [mJy]\\
            \hline
			Arp\,220\tablefootmark{a} &  0.15 & $0.65\times0.35$ &150 & 394$\pm$59\\
			Arp\,220\tablefootmark{a}&  1.4  & $0.46\times0.33$ &60 & 312$\pm$31\\
			\hline
            \noalign{\smallskip}
            \hline
         \end{tabular}
         \tablefoot{
			 \tablefoottext{a}{Integrated flux density measured by summing all
			 pixels over 3$\sigma$ associated with the Arp\,220.}
}
   \end{table}

   \begin{table}
	   \caption[]{Measured brightness at the 33\,GHz positions of the two
nuclei after convolving to the same beam of $0\farcs7\times0\farcs5$. }
         \label{tab:results}
         \begin{tabular}{ l l l}
            Frequency & East &  West\\
           {[GHz]} & {[mJy~arcsec$^{-2}$]} & {[mJy~arcsec$^{-2}$]}\\
            \noalign{\smallskip}
            \hline
			0.15 & $126\pm19\tablefoottext{a}$ & $294\pm45$\tablefoottext{a} \\
			1.4  & $425\pm43$ & $567\pm57$ \\
			6.0 & $305\pm30$ & $420\pm42$ \\
			32.5 & $111\pm11$ & $141\pm14$ \\
         \end{tabular}
         \tablefoot{
			 \tablefoottext{a}{Note that the brightness measurements at
150\,MHz may include emission not only from the nuclear disks, but may also
blend with emission from outflows and the surrounding kpc-scale disk.}}
   \end{table}

In this section we present the results obtained from the data.  The 150\,MHz
LOFAR image of Arp\,220 is presented in Fig.~\ref{fig:LOFAR}.  The 1.4\,GHz
image obtained by combining archival data from VLA and MERLIN is shown as red
contours in Fig.~\ref{fig:lofarvla}.  

\subsection{The integrated flux density at 150\,MHz}
\label{sect:discrepancy}

\begin{figure}
\centering
\includegraphics[width=0.48\textwidth]{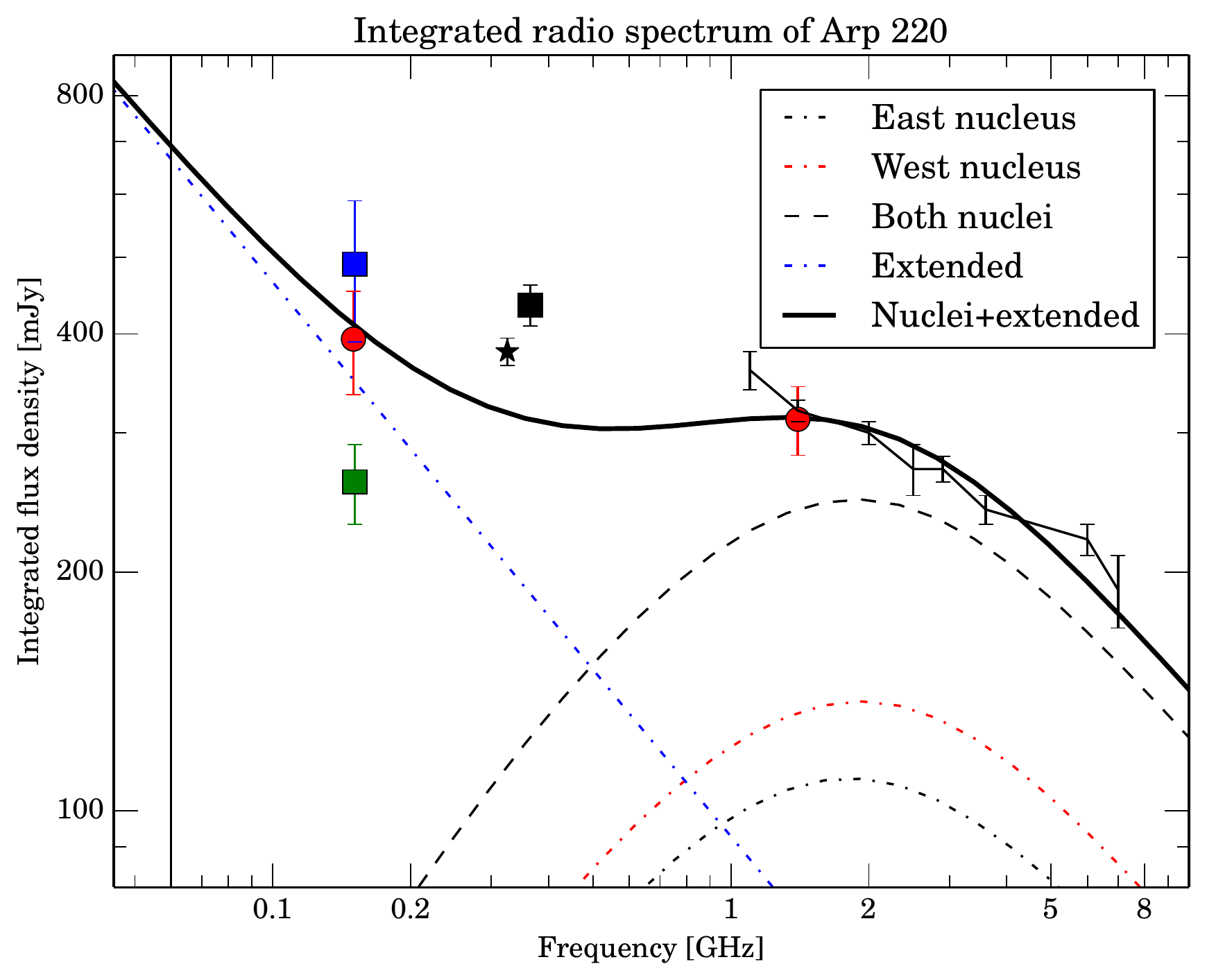}
\caption{The integrated low-frequency radio spectrum of Arp\,220. Red circles are
from this work, see Table \ref{tab:imageparams}.  Green square
(260$\pm$30@151\,MHz) from \cite{sopp1991}, blue square (490$\pm88$@151\,MHz)
from \cite{waldram1996}, black star ($380\pm15$@325\,MHz) from
\cite{anantharamaiah2000}, black square (435$\pm26$@365\,MHz) from
\cite{douglas1996}.  Black line segment from \cite{williams2010} with
resolution $250''$ at 1\,GHz to $35''$ at 7\,GHz.  See Sect.
\ref{sect:discrepancy} for a discussion on the discrepancy of the measurements
at 150\,MHz.  The dotted-dashed curves show contributions of the three
components in the model described in Sect. \ref{sect:modeling}, the dashed line
shows the sum of the two nuclei, and the solid black curve is the sum of all
three components.  The solid black vertical line indicates the frequency
60\,MHz where future observations with the International LOFAR Low Band Array
could be used to obtain $0\farcs7$ resolution images. 
}
\label{fig:spectrum}
\end{figure}

Integrated flux densities measured for Arp\,220 in this work are presented in Table
\ref{tab:imageparams}, and plotted with values from the literature in
Fig.~\ref{fig:spectrum}.  

\cite{anantharamaiah2000} and \cite{clemens2010} argue that the integrated
spectrum of Arp\,220 has a sharp turnover below 1\,GHz, although
\cite{anantharamaiah2000} find it hard to model this turnover and still fit
measurements of recombination lines. However, both these studies base their
analysis on the measurement of $260\pm30$\,mJy reported by \cite{sopp1991} at
151\,MHz using the Cambridge Low-Frequency Synthesis Telescope (CLFST) with a
beam of $250''$.  This flux density is significantly lower than what we obtain
from LOFAR.  \cite{sopp1991} state that they used flux densities from the 6C
\citep{hales1988} or 7C \citep{mcgilchrist1990} catalogues when available, and
if not, maps were made from raw CLFST data.  Since Arp\,220 is not listed in
those 6C and 7C catalogues, we assume \cite{sopp1991} made their Arp\,220 image
themselves.  
However, in the more recent 7C catalogue \citep{waldram1996} we do find a flux
density of Arp\,220 $490\pm99$\,mJy, in good agreement with our LOFAR result.
The uncertainty was calculated by us assuming 20\% flux density uncertainty for
SNR=8.3 based on \cite{waldram1996}, their Sect. 6.  The discrepancy between
\cite{sopp1991} and \cite{waldram1996} is surprising since both are using
(presumably exactly the same) CLFST data.  Using all data available in the
literature, the observations presented in this work argue that the measurement
of \cite{sopp1991} is too low and should be regarded with care when included in
interpretation or modelling.  In this paper, we exclude this data point from our
modelling.

\subsection{Spectral index maps}
\label{sect:spix}
From the continuum images at 150\,MHz, 1.4\,GHz and 6\,GHz, we made four
spectral index maps shown in Fig.  \ref{fig:spix}. In general, the nuclei
show positive spectral indices while the surrounding extended emission is 
negative or flat.

The LOFAR in-band spectral index map, Fig.~\ref{fig:spixinband}, was obtained
from the MFS-CLEAN algorithm.
We note that, as discussed in \ref{sect:fluxcal}, the spectral index of LOFAR
observations may not be reliable if the flux scale (including spectral index information)
is transferred to the target using scans where the
difference in elevation is more than 10 degrees between calibrator and target.
We emphasise that the flux calibration of these data were carefully done using
a time range where the target field and flux calibrator were close in
elevation.  Hence we are confident that the in-band spectral index map in Fig.
\ref{fig:spixinband} is reliable.

Figs.  \ref{fig:spix150_14}, \ref{fig:spix150_6} and \ref{fig:spix14_6} were
made after first convolving the respective images to the same resolution of
$0\farcs7\times0\farcs5$, position angle 161$^\circ$, before calculating
$\alpha$ pixel-wise (with a pixel size of $0\farcs02$). i
The convolved images had root-mean-square noise levels of
$RMS_{150\,MHz}=164\,\mu$Jy\,beam$^{-1}$,
$RMS_{1.4\,GHz}=93\,\mu$Jy\,beam$^{-1}$, $RMS_{6.0\,GHz}=20\,\mu$Jy\,beam$^{-1}$.
The spectral index maps were calculated as $\alpha_{\nu1,2}=\log(S_2/S_1)/\log(\nu_2/\nu_1)$.

\subsubsection{Spectral index error estimates}
\label{sect:errspix}
The LOFAR in-band spectral index image was obtained through the MFS-algorithm,
which also produced pixel-wise error-estimates for the spectral index (Fig.
\ref{fig:errspixinband}).  

For the other three spectral index maps in Fig.
\ref{fig:spix}, we estimated the uncertainty of the spectral index, accounting for both the image
noise and possible systematic offsets in the flux scale, as done by e.g.
\cite{kim2014}, their Sect. 2.5. We first define the total intensity error (per
pixel) as
\begin{equation}
\sigma_\nu = \delta_\nu S_\nu + RMS_\nu
\end{equation}
where RMS$_\nu$ is noise of the convolved images, $S_\nu$ is the measured
surface brightness, and $\delta_\nu$ is the absolute flux density uncertainty
(i.e. 15\% at	150\,MHz and 10\% at 1.4\,GHz and 6\,GHz).  The (pixel-wise)
uncertainty of the spectral index is calculated via standard error propagation
as
\begin{equation}
E(\alpha_{\nu1,2})= \frac{1}{\log(\nu_2/\nu_1)}\times\left[\frac{\sigma_{\nu1}^2}{S_{\nu1}^2}+\frac{\sigma_{\nu2}^2}{S_{\nu2}^2}\right]
\end{equation}
The resulting three error maps can be seen in Figs. \ref{fig:errspix150_14},
\ref{fig:errspix150_6}, and \ref{fig:errspix14_6}.

	\subsubsection{Clipping of spectral index maps}
\label{sect:clipping}
\begin{figure*}[htbp]
\centering
\subfigure[LOFAR in-band $\alpha$, 127-174\,MHz]{
	\includegraphics[width=0.48\textwidth]{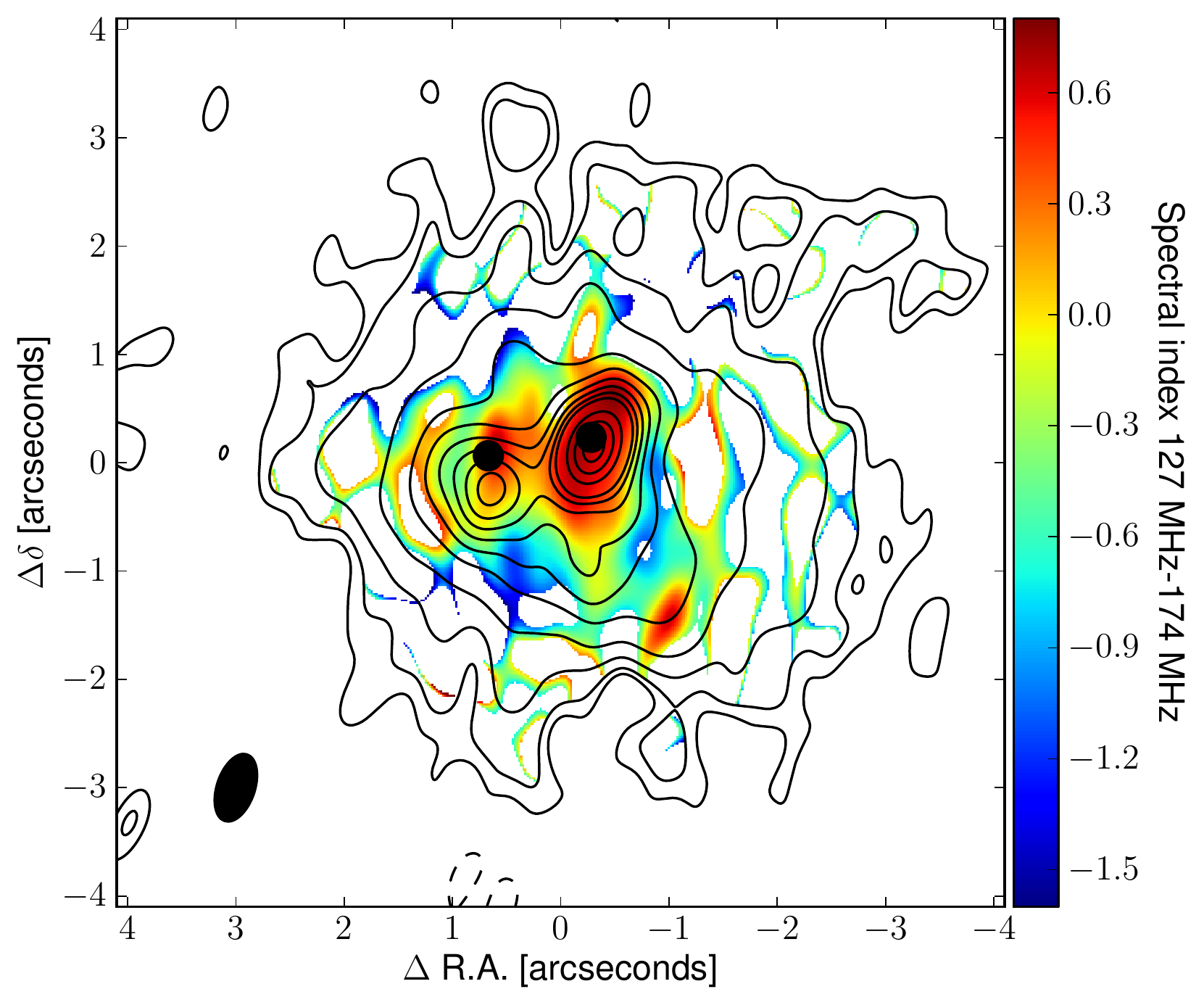}
        \label{fig:spixinband}
}
\subfigure[150\,MHz-1.4\,GHz]{
	\includegraphics[width=0.48\textwidth]{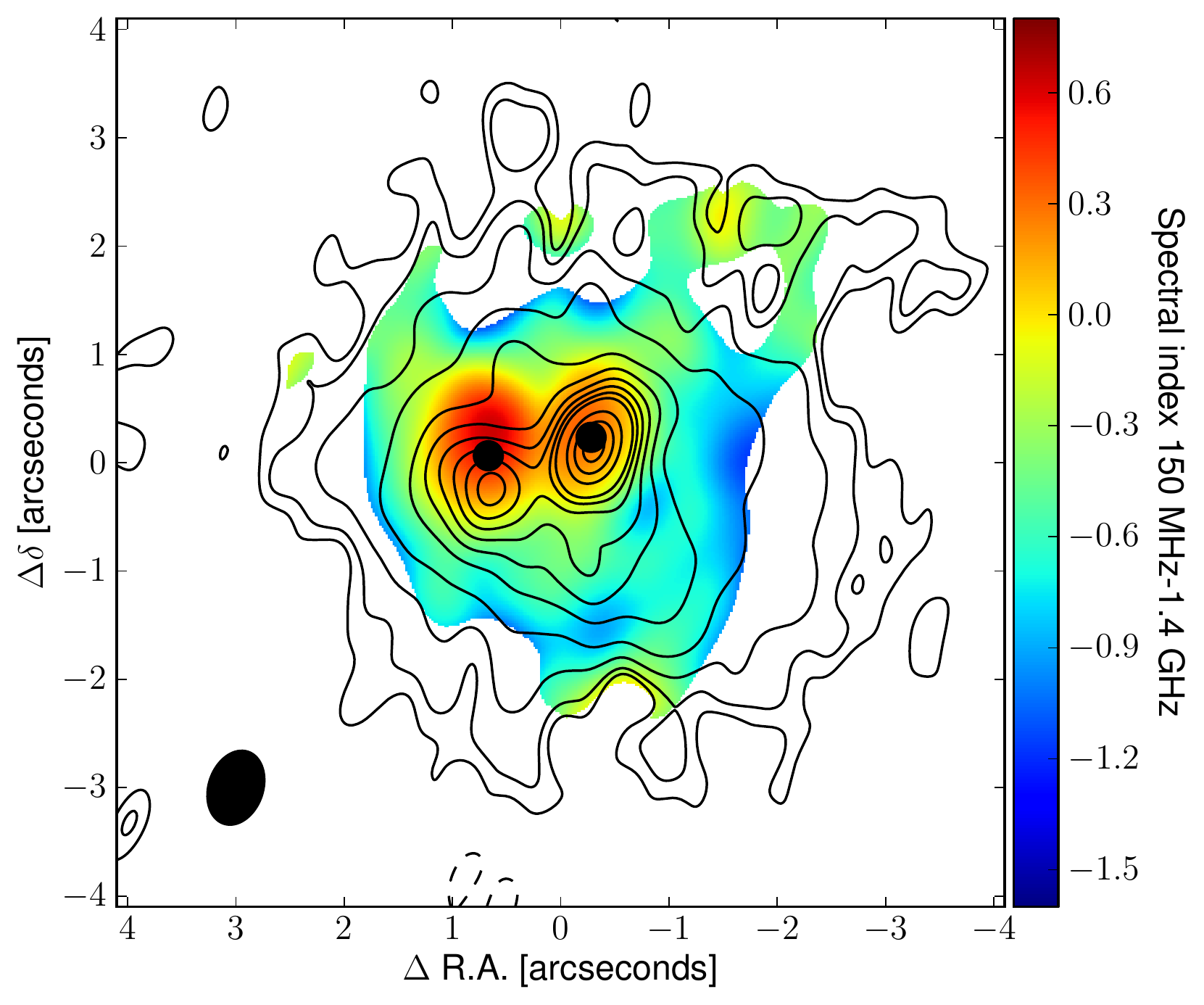}
        \label{fig:spix150_14}
}
\subfigure[150\,MHz-6\,GHz]{
	\includegraphics[width=0.48\textwidth]{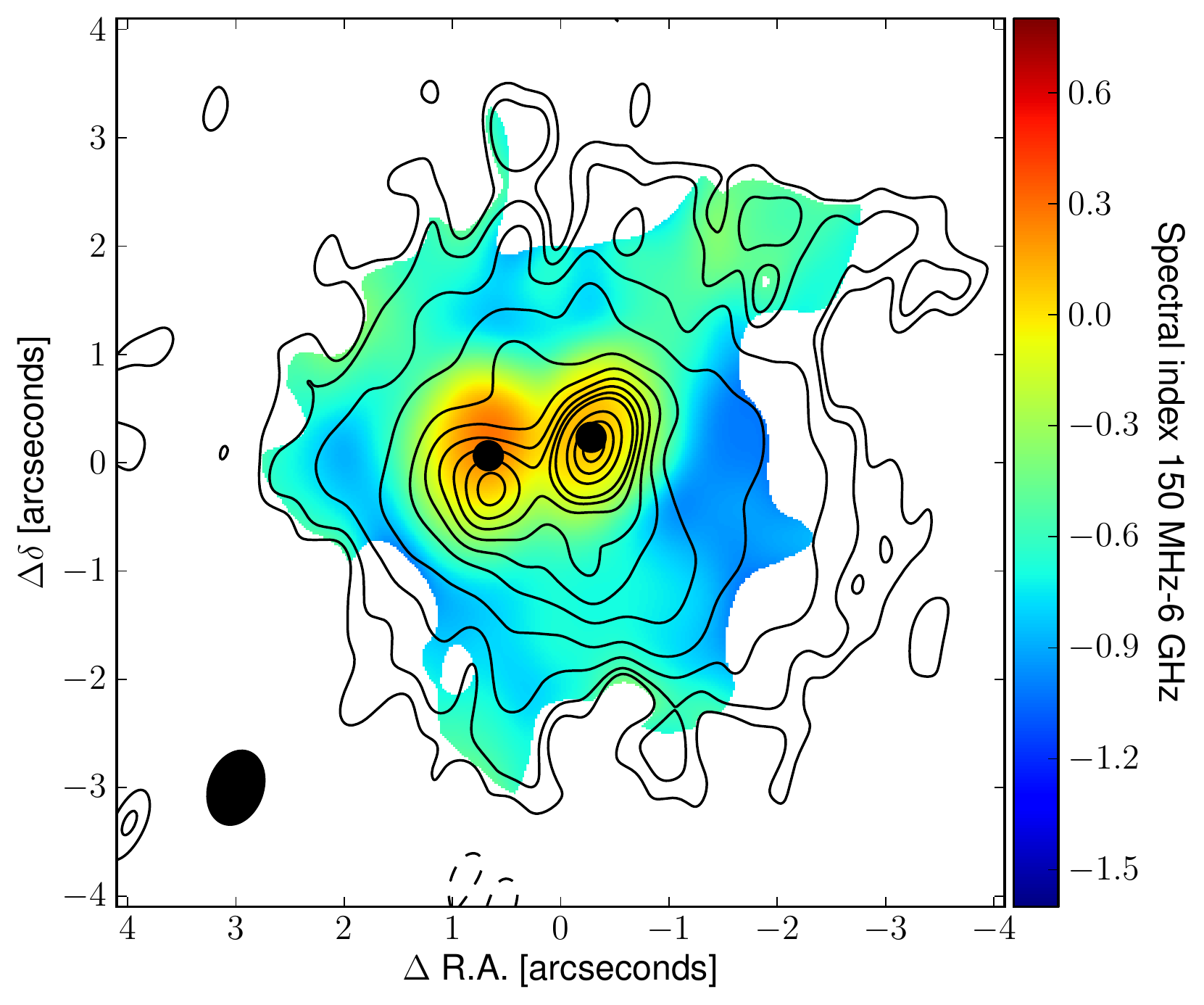}
        \label{fig:spix150_6}
}
\subfigure[1.4\,GHz-6\,GHz]{
	\includegraphics[width=0.48\textwidth]{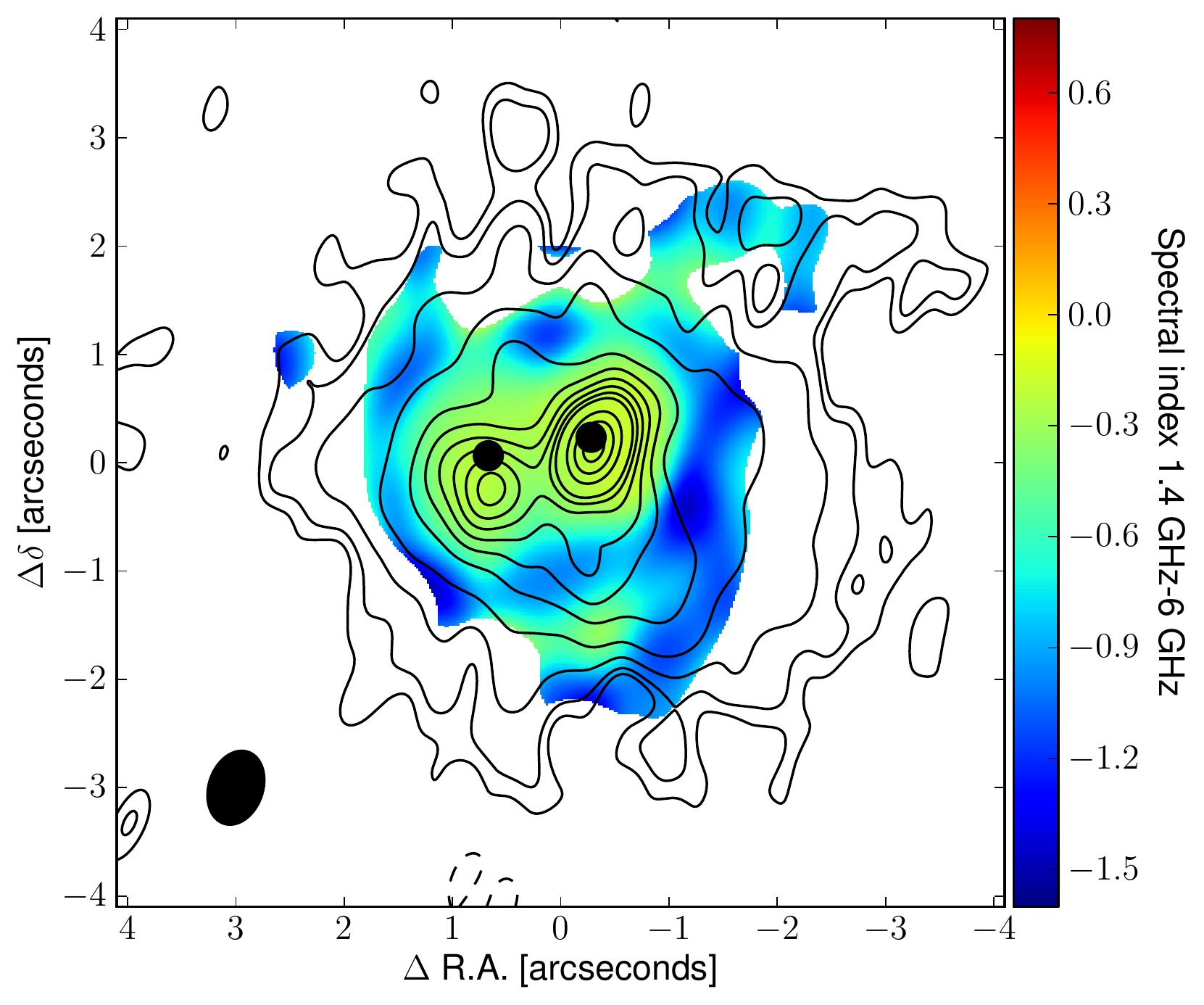}
        \label{fig:spix14_6}
}
\caption{Spectral index maps using data from three radio bands, overlayed with
	the same 150\,MHz contours as in Fig.~\ref{fig:LOFAR}.  Panel
	\subref{fig:spixinband} shows the in-band spectral index as recovered by
	MFS-CLEAN of the LOFAR data covering 127-174\,MHz.  Panel
	\subref{fig:spix150_14} shows the spectral index between 150\,MHz and
	1.4\,GHz, panel \subref{fig:spix150_6} between 150\,MHz and 6\,GHz, and
	panel \subref{fig:spix14_6} between 1.4\,GHz and 6\,GHz.
	The (convolved) resolution is plotted in the
	lower left of each panel. The black dots in the centre mark the 33\,GHz
positions as in Fig.~\ref{fig:LOFAR}.  White pixels were clipped as described in Sect.
\ref{sect:clipping}.  \label{fig:spix}
}
\end{figure*}

\begin{figure*}[htbp]
\centering
\subfigure[LOFAR in-band $\alpha$, 127-174\,MHz]{
	\includegraphics[width=0.48\textwidth]{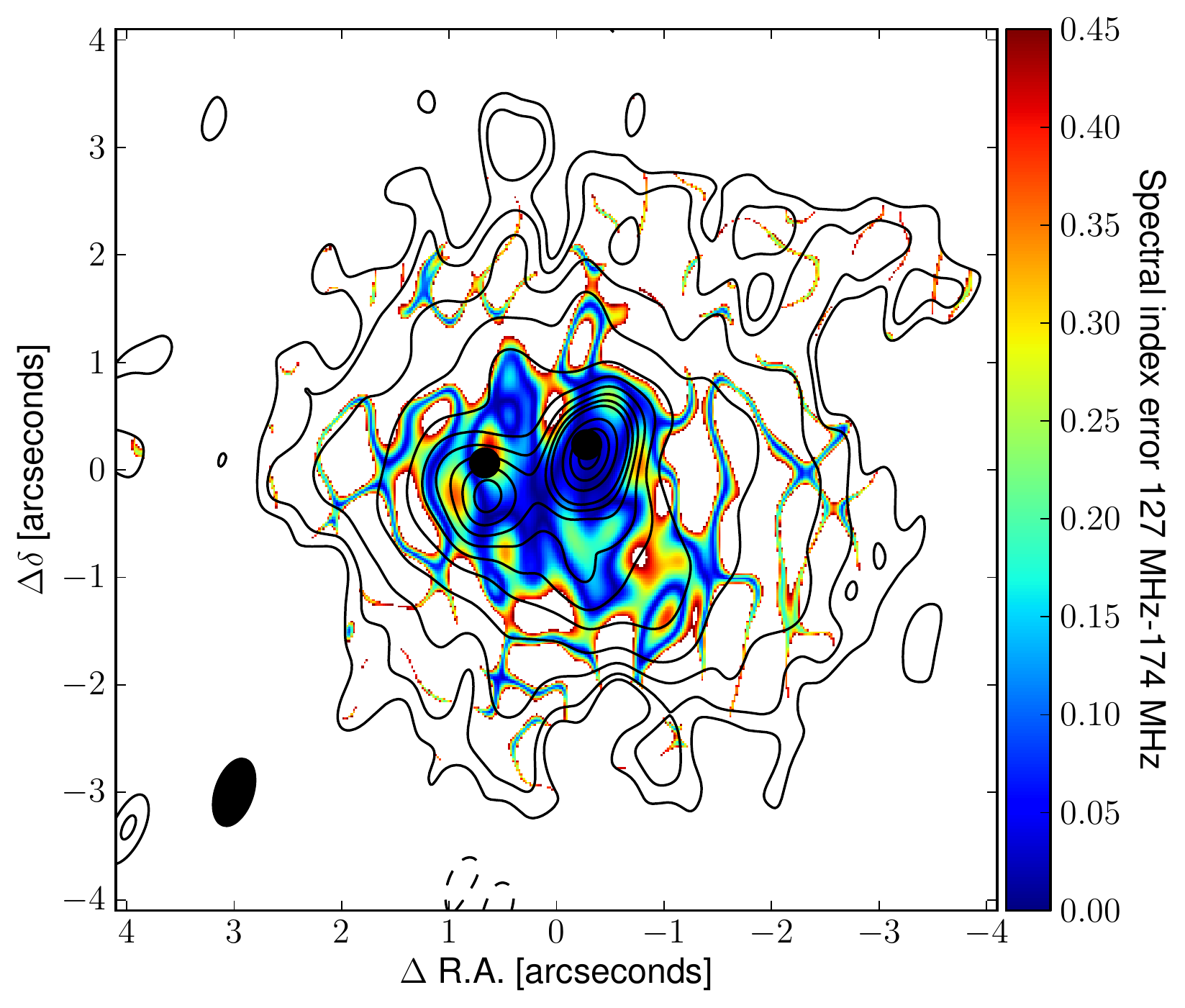}
        \label{fig:errspixinband}
}
\subfigure[150\,MHz-1.4\,GHz]{
	\includegraphics[width=0.48\textwidth]{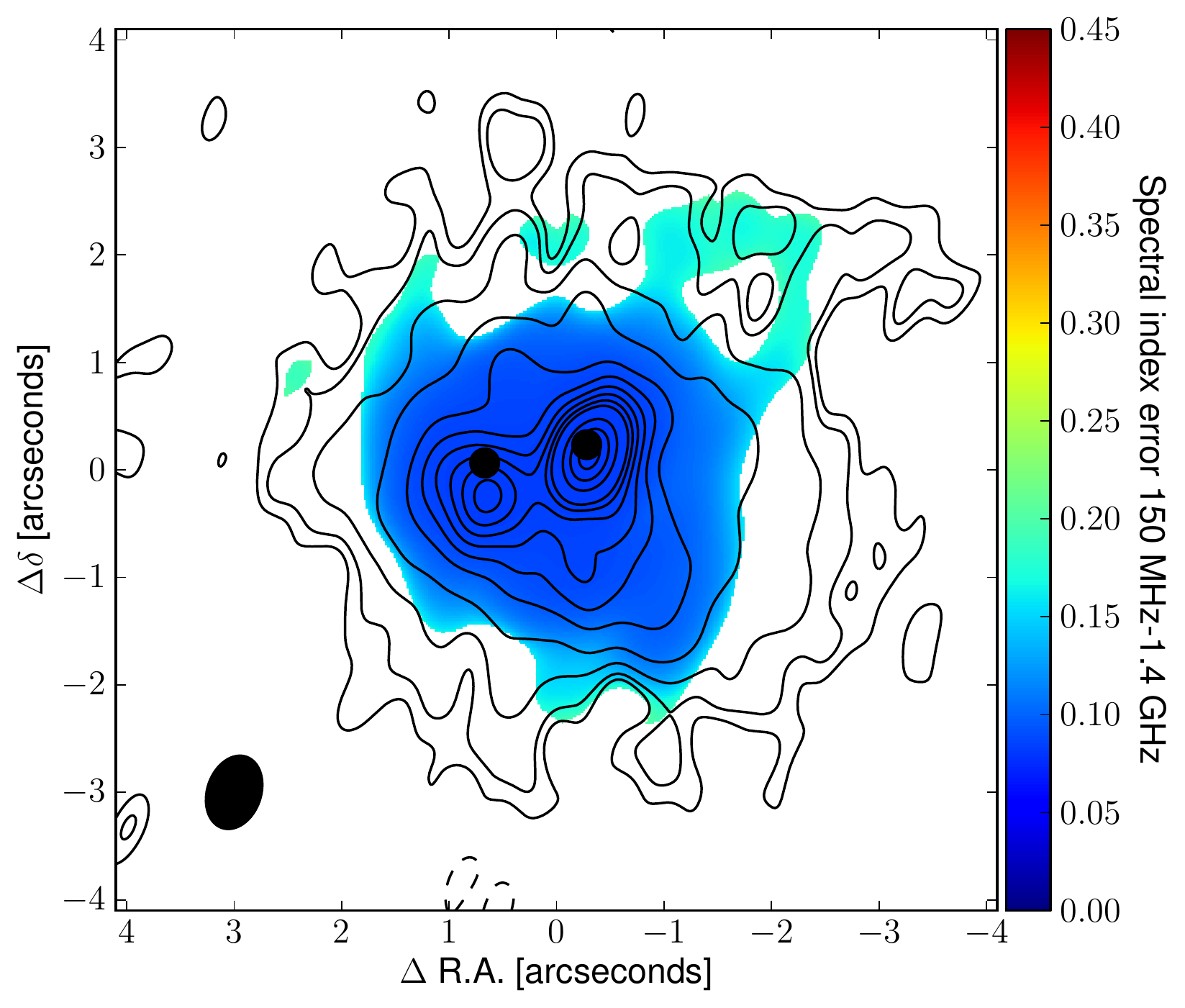}
        \label{fig:errspix150_14}
}
\subfigure[150\,MHz-6\,GHz]{
	\includegraphics[width=0.48\textwidth]{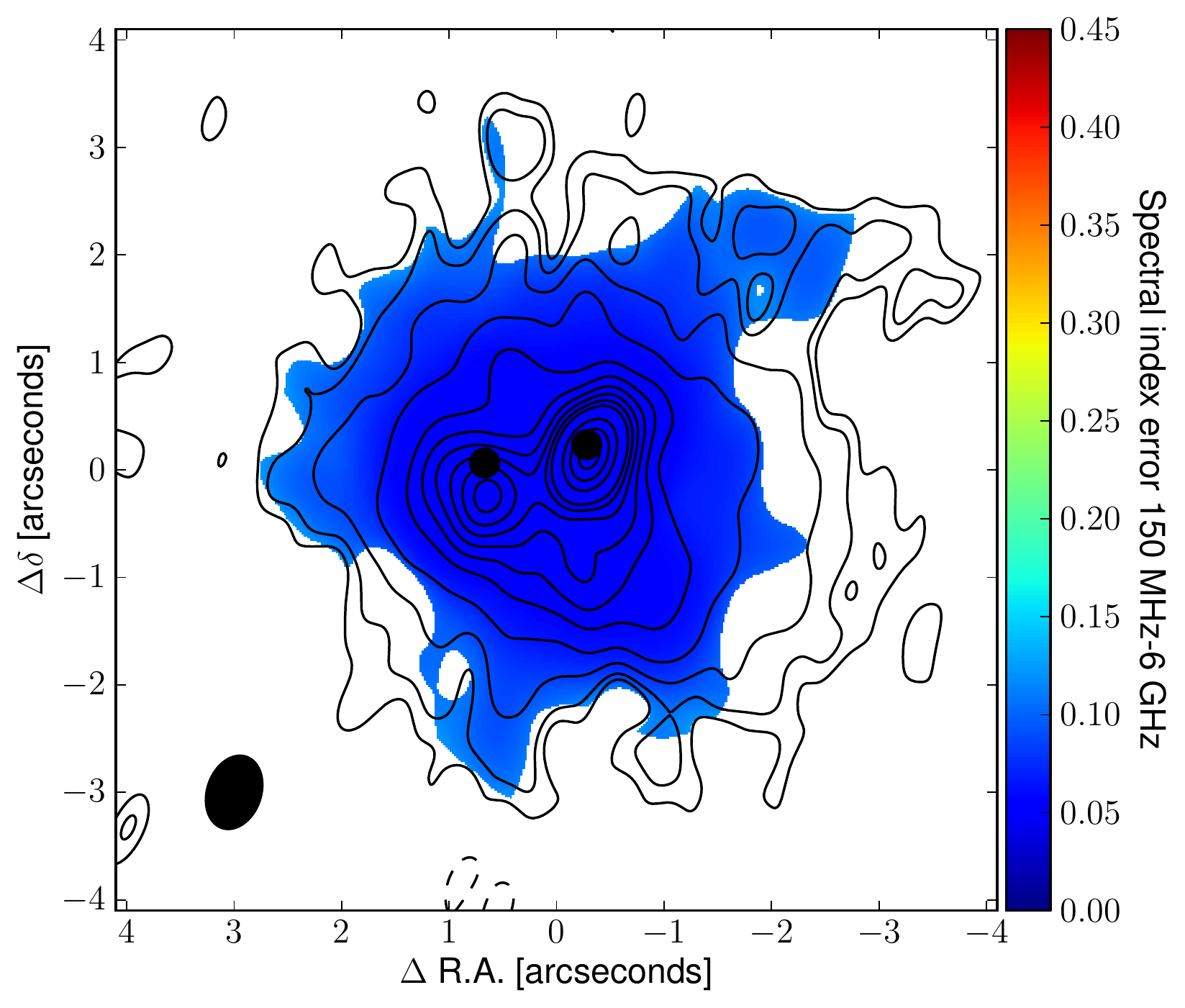}
        \label{fig:errspix150_6}
}
\subfigure[1.4\,GHz-6\,GHz]{
	\includegraphics[width=0.48\textwidth]{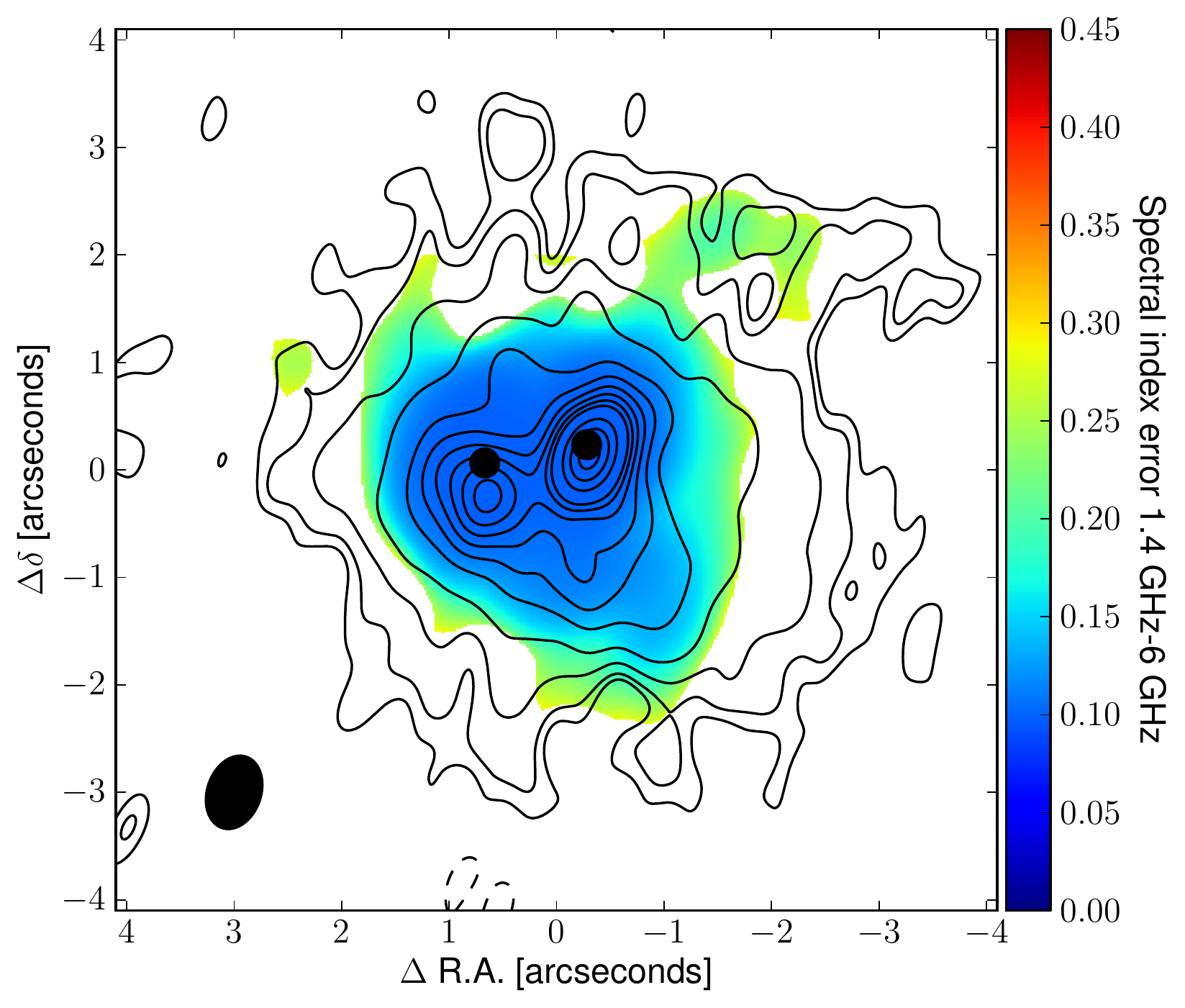}
        \label{fig:errspix14_6}
}
\caption{Uncertainty estimates, obtained as described in Sect. \ref{sect:errspix},
for the spectral index maps presented in Fig. \ref{fig:spix}.
	The (convolved) resolution is plotted in the
	lower left of each panel. The black dots in the centre mark the 33\,GHz
positions as in Fig.~\ref{fig:LOFAR}.  White pixels were clipped as described in Sect.
\ref{sect:clipping}.  \label{fig:errspix}
}
\end{figure*}

In some regions of the images the spectral index uncertainties are large, in particular
in the weakest parts where the signal-to-noise ratio is low. 
	To ensure reliable spectral index maps, we therefore 
	conservatively blanked (white colour) Fig.~\ref{fig:spix} and
	Fig.~\ref{fig:errspix} where either of the (convolved)
	input continuum images were weaker than $5\sigma$. 
	However, for the LOFAR in-band spectral index we noted significant uncertainties
	also in regions with brighter emission. 
	We therefore blanked Figs. \ref{fig:spixinband}
	and \ref{fig:errspixinband} where the uncertainty in the spectral index (Fig. \ref{fig:errspixinband}) was
	larger than 0.45. 
	We note that the MFS algorithm with two Taylor terms cannot 
fully represent any spectral curvature within the LOFAR band which could be present due to
free-free absorption of emission close to the nuclei. This could be a reason for the 
large uncertainties reported by the MFS algorithm in some regions of this image.

	\subsubsection{Sampling of extended emission}
	We are confident that we are not missing any significant large-scale
	emission which could affect our spectral index measurements.  Assuming the
	largest angular scale sampled in the observations can be estimated as
	$\lambda/b$ radians where $\lambda$ is the observing wavelength and $b$ is
	the shortest baseline included in imaging, we estimate the largest scales
	sampled as $1.7'$, $1.9'$ and $4.9'$ at 150\,MHz, 1.4\,GHz and 6\,GHz
	respectively. This is much larger than the extent of Arp\,220, which covers
	at most $5''$ in any of the images.


\subsection{The eastern nucleus}
In Fig.~\ref{fig:LOFAR} we measure a peak brightness associated with the eastern
nucleus of $22.0\pm3.3$\,mJy~beam$^{-1}$ at R.A.  $15^{\rm h}34^{\rm m}57^{\rm s}.289$, Dec.
$23^\circ30'11\farcs04 $, i.e. a significant offset of 300\,mas (110\,pc)
\emph{south} of the position reported for the eastern nucleus at 33\,GHz by
\cite{barcosmunoz2015}.  However, in Fig.~\ref{fig:spixinband}, which is made
from the same data, we detect a feature of positive spectral index with a peak
100\,mas \emph{north} of the 33\,GHz position. A closer look reveals a similar
spectral feature in Figs. \ref{fig:spix150_14} and \ref{fig:spix150_6}.  
At the GHz position of the eastern nucleus, the low 150\,MHz surface brightness
and the positive spectral index indicates significant absorption of the
emission from the nuclear disk. In fact, the emission measured at 150\,MHz at
this position may be due to the southern component (eastern 150\,MHz peak)
blending with the GHz position due to the finite image resolution, and hence
our measurements are consistent with the eastern nuclear disk being completely
absorbed at 150\,MHz.

\subsection{The western nucleus}
The western nucleus is the brightest component at 150\,MHz.  In
Fig.~\ref{fig:LOFAR} we measure a peak brightness associated with the western
nucleus of $48.8\pm7.3$\,mJy~beam$^{-1}$ at R.A.  $15^{\rm h}34^{\rm m}57^{\rm
s}.219$, Dec.  $23^\circ30'11\farcs44$, i.e. 70\,mas (25\,pc) south and 40\,mas
west of the 33\,GHz position reported for the western nucleus by
\cite{barcosmunoz2015}, although the offset is not significant given the
astrometric uncertainties of the 150\,MHz and 33\,GHz data.  However, the
150\,MHz contours at levels of [40,60,80]$\times \sigma$ in Fig.
\ref{fig:LOFAR}, show an extension $\sim1''$
(330\,pc) south of the western nucleus.  Indeed, we see a similar extension at
1.4\,GHz and 6\,GHz (Figs.  \ref{fig:lofarmerlin} and \ref{fig:lofarvla}).  We
also find a north-south elongated feature of positive spectral index in the
LOFAR in-band spectral index map (Fig.~\ref{fig:spixinband}).

\subsection{The extended emission}
The structure of the extended emission detected with LOFAR at 150\,MHz matches
very well the structure 1.4\,GHz (Fig.  \ref{fig:lofarmerlin}) and 6\,GHz
(Fig.~\ref{fig:lofarvla}). At 150\,MHz the extended emission accounts for more
than 80\% of the total flux density of Arp\,220.

Although Fig.~\ref{fig:spixinband} only covers a small area around the nuclei,
the extended emission is consistent with having a flat or negative spectral
index.  The spectral index maps between 150\,MHz and 1.4\,GHz
(Fig.~\ref{fig:spix150_14}), and between 150\,MHz and 6\,GHz
(Fig.~\ref{fig:spix14_6}), show that the extended emission has a spectrum of
$\alpha\approx-0.7$ down to 150\,MHz, consistent with optically thin
synchrotron emission.  Fig.~\ref{fig:spix14_6} shows a similar overall spectral
index for the extended emission, although with steeper values towards the
edges. This could be evidence of synchrotron ageing, but because of the smaller
separation in frequency compared to Figs.~\ref{fig:spix150_14} and
\ref{fig:spix14_6}, as well as potentially low signal to noise in the outer
regions, we do not include this in our modelling.

\section{Modelling}
\label{sect:modeling}
Since we now resolve Arp\,220 from 150\,MHz to 33\,GHz, we can model the
spectrum of Arp\,220 pixel-by-pixel. From Fig.~\ref{fig:LOFAR} it is clear that
at least three components have to be considered: the two nuclei and the
extended emission. 

\subsection{Modelling the nuclei}
\label{sect:nucleimodel}
For the structure of the nuclei we use the models fitted by
\cite{barcosmunoz2015} at 33\,GHz, where each nucleus is a geometrically thin
(but optically thick) inclined exponential disk.  The surface brightness of
each disk is described as
\begin{equation}
	S(x,y) = A\times\text{exp}\left(-\frac{\sqrt{x^2 + (y/\cos i)^2}}{l}\right)
\end{equation}
where $S(x,y)$ is the surface brightness at position (x,y)
relative to the centre of the disk (with some P.A.), $A$ is the peak
brightness, $i$ is the inclination angle of the disk, and $l$ is a
characteristic scale length. Using the deconvolved values reported by
\cite{barcosmunoz2015}, their Table 2 (East: A=6.0\,mJy~beam$^{-1}$,
P.A.=54.7$^\circ$, $l$=30.3\,pc, $i$=57.9$^\circ$, West:
A=13.4\,mJy~beam$^{-1}$, P.A.=79.4$^\circ$, $l$=21.0\,pc, $i$=53.5$^\circ$), we
obtain a model with total flux densities of 23.4\,mJy and 28.1\,mJy at 33\,GHz for the
east and west nuclei respectively.  However, with these values the
model systematically underpredicts the measured surface brightness at all
frequencies (including 33\,GHz) at the 33\,GHz positions of the nuclei (after we
convolve the model and images to a common lower resolution for a pixel-wise
comparison). We therefore scale the model to 26.7\,mJy and 31.1\,mJy for the
east and west nuclei respectively to match the data at 33\,GHz. We note that
these higher values are within the uncertainties of the measurements of
$30.1\pm3.9$\,mJy and $33.4\pm4.0$\,mJy reported by \cite{barcosmunoz2015},
their Table 1. 

To model the nuclei, we assume that each pixel follows the model presented by
\cite{condon1991} for radio emission from a star forming galaxy. In this model
the emitting medium is a well-mixed thermal/non-thermal plasma, and the radio
spectrum is given by the expression
\begin{equation} 
    S_\nu=
    10^{-1.3}\left(\frac{\nu}{8.4\mathrm{~GHz}}\right)^2  T_e
    \left(1-e^{-\tau_\nu}\right)
    \left[1+f_\text{th}^{-1}\left(\frac{\nu}{1\mathrm{~GHz}}\right)^{\alpha+0.1}\right],
    \label{eqn:condon}
\end{equation}
where $S_\nu$ is the surface brightness in mJy~arcsec$^{-2}$, $T_e=7500$~K is
the thermal electron temperature \citep{anantharamaiah2000}, $\alpha$ is the
synchrotron radiation spectral index \citep{condon1992}, $f_\text{th}$ is the
fraction of thermal (free-free) emission at 1\,GHz, and
$\tau_\nu=(\nu/\nu_c)^{-2.1}$ is the free-free optical depth expressed in terms
of a turnover frequency $\nu_c$ where $\tau_\nu=1$. Given $S_\nu$ from the
exponential disk model at 32.5\,GHz, and assuming values for $f_\text{th}$ and
$\alpha$ for each nuclei, we can calculate $\nu_c$ (for each pixel) and thereby
predict the radio spectrum for the nuclei from Eq. \ref{eqn:condon}.

\subsection{Modelling the extended emission}
\label{sect:sphere}
Although it is clear from Fig.~\ref{fig:LOFAR} that the extended emission has 
structure, we model it simply as an optically thin
sphere of uniform density and radius $r_S$.  Projected in two dimensions, the
emission at radius $r<r_S$ is described by
$S\propto2r_S\sin(a), \text{ where } a = \arccos(r/r_S)$.
The flux scale is fixed by normalising the sum of all pixels to a chosen total
integrated flux density of the sphere.
We fix the centre position of the sphere to the peak position of CO(1-0) in
Fig.~\ref{fig:co10}, taken to be at R.A. $15^{\rm h}34^{\rm m}57^{\rm s}.240$,
Dec.  $23^\circ30'11\farcs22$, and the radius to $2''$.  Although the
extended emission shows a range of spectral indices, from about 0 (closer to
the nuclei) to $-1$ in Fig.~\ref{fig:spix}, the major part should be well
described by a typical synchrotron spectral index of $\approx-0.8$
\citep{condon1992}.  To account for free-free absorption in the brightest
regions close to the nuclei, we assume a marginally flatter spectral index of
$-0.7$ for the sphere, in agreement with the overall value reported in Sect.
\ref{sect:results}.

\subsection{Modelling results}
\label{sect:modelingres}
\begin{figure*}
\centering
\includegraphics[width=\textwidth]{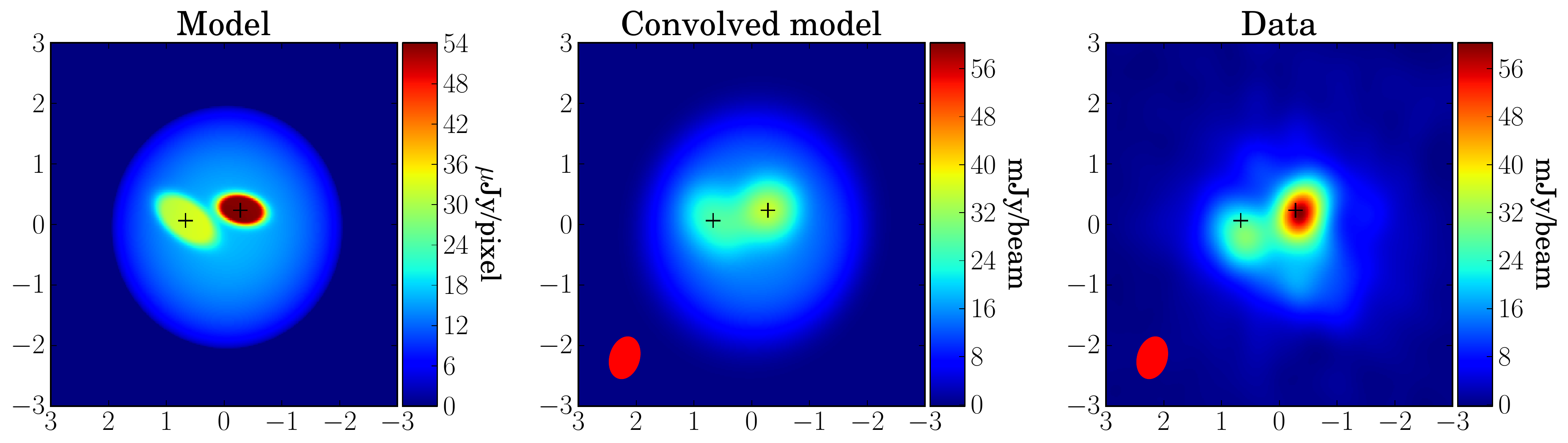}
\caption{A three-component model of two exponential disks and a uniform sphere
as described in Sect. \ref{sect:modeling}. The left panel shows the model
prediction at 150\,MHz. The middle panel shows the same model after convolving
to a resolution of $0\farcs7\times0\farcs5$ (shown as a red ellipse in the
lower left). The right panel shows the data at 150\,MHz convolved to the same
resolution. Note that the middle and right panels have the same colour scale for
easy comparison. The crosses mark the 33\,GHz positions of the two nuclei.
Note that the nuclei follow Eq.  \ref{eqn:condon} and are therefore almost
completely saturated at 150\,MHz.
}
\label{fig:model}
\end{figure*}

By adding the models of the nuclei and extended emission we can now predict the
resolved spectrum of Arp\,220 at any (radio) frequency. This three-component
model is shown in Fig.~\ref{fig:model}. 

The sphere, described in Sect. \ref{sect:sphere}, was modelled with a total flux
density of 8.1\,mJy at 32.5\,GHz and spectral index $-0.7$ (i.e. 350\,mJy at
150\,MHz). The exponential disks, described in Sect. \ref{sect:nucleimodel},
were modelled with flux densities of 26.7\,mJy (east) and 31.1\,mJy (west) at
32.5\,GHz, non-thermal spectral indices ($\alpha$ in Eq. \ref{eqn:condon}) of
$-0.65$ (east) and $-0.70$ (west), and thermal fractions at 1\,GHz ($f_\text{th}$ in
Eq. \ref{eqn:condon}) of 0.8\% (east) and 0.4\% (west).

\begin{figure}
\centering
\includegraphics[width=0.48\textwidth]{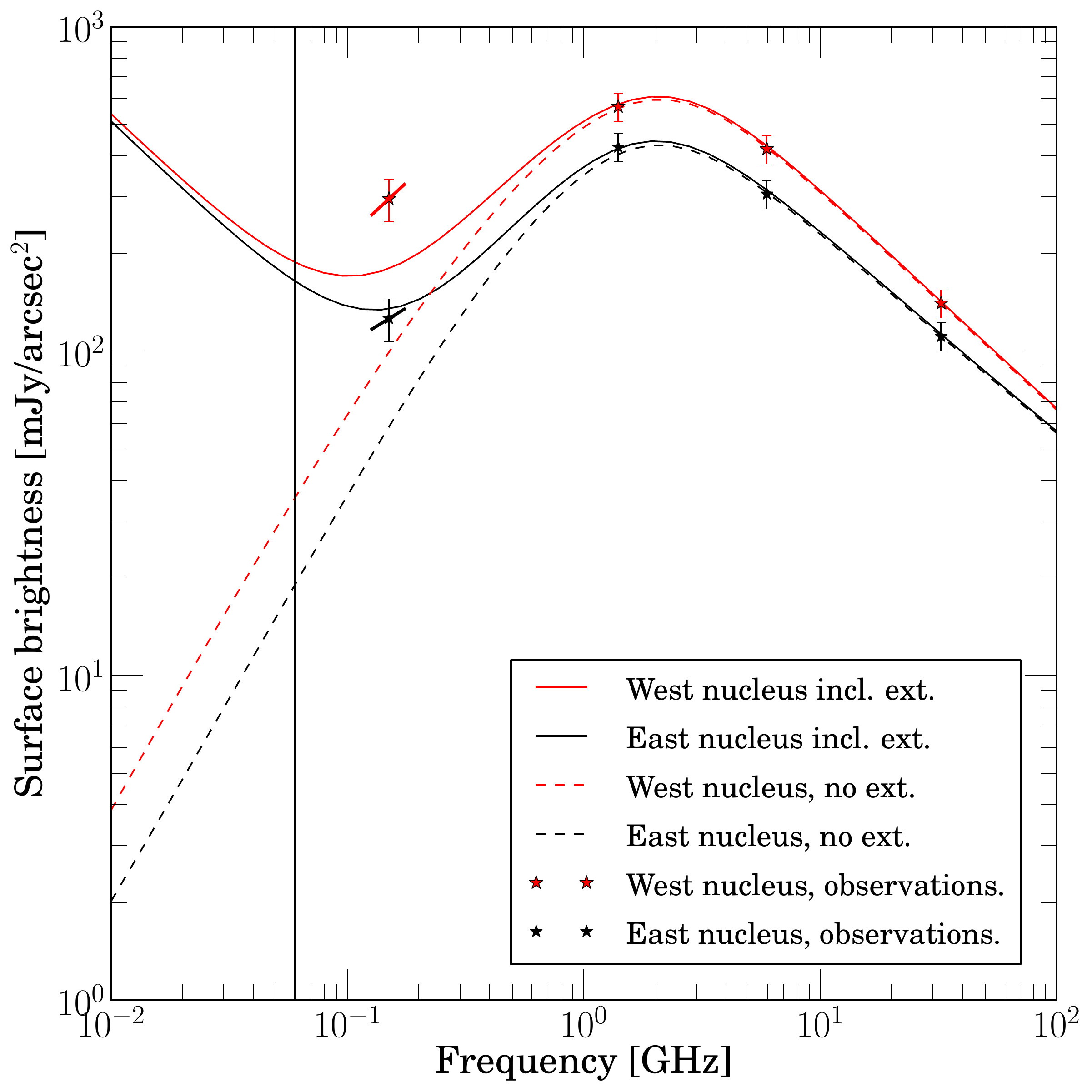}
\caption{Measured and modelled spectra of the eastern (black) and western (red)
	nuclei.  Surface brightness measurements (Table \ref{tab:results}) are
	plotted as stars. The thick line segments going through the 150\,MHz
	measurements illustrate the measured LOFAR in-band spectral index at the
GHz positions of the nuclei, see also Fig.~\ref{fig:spixinband}.  
	The solid lines show the surface brightness at 33\,GHz positions of the two nuclei,
	as predicted by the three component model described in Sect.
	\ref{sect:modeling}.
	The dashed lines show the spectra of the nuclei without any extended component,
i.e. only the exponential disks described in Sect. \ref{sect:nucleimodel} following Eq. \ref{eqn:condon}.
	The thin black vertical line indicates the frequency 60\,MHz where future
observations with the International LOFAR Low Band Array could be used to
obtain $0\farcs7$ resolution images.}
\label{fig:nuclei}
\end{figure}

To compare the model with measurements at various frequencies, we measure the
integrated flux density of Arp\,220, as well as the surface brightness at the
GHz positions of the two nuclei.  The total flux density is measured by summing
the pixels in the model at each frequency. This can directly be compared with
the measured integrated flux density across the radio spectrum. 
To measure the surface brightness of the nuclei it is important to take into
account the smoothing caused by the finite resolution of the images.
Therefore, we convolve the model and data at each frequency to the same
resolution of $0\farcs7\times0\farcs5$, position angle 161$^\circ$, before
measuring the brightness of the nuclei.  The measured surface brightness values
for the nuclei are presented in Table \ref{tab:results} and plotted in
Fig.~\ref{fig:nuclei}.  A comparison of the model and the data is shown at
150\,MHz in Fig.~\ref{fig:model}.

\subsubsection{The integrated spectrum}
We find that our model can reproduce the integrated flux density measured
through most of the radio spectrum, see the solid line in Fig.
\ref{fig:spectrum}.  However, we note that the model underpredicts the
flux density of the Arp\,220 in the range 200\,MHz to 1\,GHz.

Although the structure of the extended emission is clearly not a perfect
sphere, the predicted surface brightness is close to what is observed at
150\,MHz. 

We note that at very low frequencies the spectrum must flatten, even
without thermal absorption, because of e.g. the increased importance
of synchrotron self-absorption in regions of very high surface brightness.
This could be investigated by international LOFAR observations using the Low
Band Array (LBA), which could obtain a matched synthesised resolution
$0\farcs7$ at 60\,MHz.  Such observations are currently very challenging, but
commissioning work is ongoing to improve the capabilities. 

\subsubsection{The nuclei}
Turning to the brightness of each nucleus, we find that our model correctly
predicts the western nucleus to be brighter at all frequencies, and that both
spectra turn over around 1\,GHz. Above 1\,GHz the spectrum is well described by
the two nuclei following Eq.~\ref{eqn:condon}.  We find that including the
extended emission, which in our model is represented by parts of the sphere
close to the nuclei, makes the two nuclei significantly brighter at 150\,MHz
(solid lines in Fig.~\ref{fig:nuclei}) than without the extended emission
(dashed lines in Fig.~\ref{fig:nuclei}). We also find that the model with 
extended emission better reproduces the spectral indices measured for the two
nuclei within the LOFAR band, shown as thick solid line segments going through
the 150\,MHz measurements in Fig.~\ref{fig:nuclei}. 

From our modelling in Sect.~\ref{sect:modeling} it is clear that the nuclei can
be described by well mixed thermal/non-thermal exponential disks at GHz
frequencies.  This argues against that the nuclei are clumpy \citep{lacki2013},
in agreement with recent ALMA observations presented by \cite{scoville2015}.

However, at 150\,MHz, the model underpredicts the observed surface brightness
of the western nucleus, even when taking into account the extended emission,
predicting 180\,mJy~arcsec$^{-1}$ (after convolving to
$0\farcs7\times0\farcs5$), i.e. 60\% of the observed value.  This indicates
that there is emission close to this nucleus which is not included in the
model. Indeed, outflows are not included in our simple model, and may very
well be responsible for the extra emission detected from the western nucleus.
This emission may also partly explain why the model underpredicts the
integrated flux density in the range 200\,MHz to 1\,GHz.

\section{Discussion}
\label{sect:discussion}
In this section we discuss the results of observations and modelling and compare 
with literature. 

\subsection{Low thermal fractions in the nuclei}
The thermal fractions at 1\,GHz, required to match the data, are only 0.8\% and
0.4\% for the east and west nuclei respectively. This is an order of magnitude
lower than the $9\%\pm3\%$ found by \cite{marvil2015} for a large sample of
galaxies, although \cite{murphy2013} finds lower thermal fractions for ongoing
mergers.  We note that our modelling assumes no foreground thermal absorption of
the nuclei, although this could also affect the synchrotron emission from the
nuclei and therefore the thermal fractions reported here should be considered
upper limits.  As noted by \cite{barcosmunoz2015} based on the lack of thermal
emission at 33\,GHz, the easiest explanation for the lower-than-expected
thermal flux is that a significant fraction of the ionising photons produced by
young stars is absorbed by dust. 

\subsection{An outflow in the eastern nucleus}
\label{sect:eastern}
CO(2-1) observations of the eastern nucleus suggests it to be a rotating disk
seen almost edge on \citep{sakamoto2008,koenig2012}. In Fig.~\ref{fig:co21} we
show the structure of CO(2-1) in the nuclei, showing the orientation of the
eastern nucleus.  The 150\,MHz features detected north and south of the eastern
nuclear disk (Sect. \ref{sect:results}) are consistent with an elongated
structure where the northern part is viewed through a free-free absorbing
medium, possibly the outer part of the nuclear disk, and the southern part is
relatively unaffected by this absorbing medium.  The fact that the features are
on opposite sides of the centre of the nuclear disk indicates that the emission is
associated with the disk.  

\cite{sakamoto2009} presents evidence for outflows in the nuclei with speeds of
100~km~s$^{-1}$ based on P Cygni profiles in HCO$^+$ and CO.  We expect that
this outflow could carry radio emitting CRs from the nucleus, or accelerate
CRs in situ in the outflow due to shocks and turbulence. This would manifest
itself as synchrotron emission tracing an elongated or bipolar feature extending
outwards from the centre of the disk, similar to what we observe at 150\,MHz.
We note, however, that \cite{barcosmunoz2015} fit a P.A. of 
54.7$^\circ$ for the eastern disk, and a perpendicular outflow would then have
P.A. = -35.3$^\circ$.  The elongation we observe implies a P.A.$\sim0^\circ$,
i.e. the 150\,MHz feature is not perfectly aligned with the disk. 

\emph{HST} NICMOS images of Arp\,220 presented by \cite{scoville1998} show a
north-south extension of the eastern nucleus at 1.6\,$\mu$m and 2.2\,$\mu$m.
The nature of this structure is not clear, but \cite{scoville1998} argue, based
on the extinction structure, that the south side is the near side.
Similar north-south structure is also seen in the 3.8\,$\mu$m VLT images
presented by \cite{gratadour2005}, their Fig. 5.

Another galaxy with a bright star forming disk seen almost edge-on is M82. This
galaxy was recently found to have bright radio continuum emission at 150\,MHz
around its star forming disk \citep{varenius2015}. \cite{varenius2015}
interpret this as the base of the outflow seen at larger distances from M82.
From their figure 3b (made at 154\,MHz) it is clear that the emission is
brighter on the south-east side of M82, presumably because it is closer and
less obscured by free-free absorption in the disk which is partly overlapping
the north-west emission. The outflow seen in M82 is brightest about 170\,pc
(10$''$) from the centre of the star forming disk, i.e. similar to the 110\,pc
(0$\farcs3$) offset in the eastern nucleus of Arp\,220 between the peaks at
33\,GHz and 150\,MHz.  

Based on the consistency with previous evidence for an outflow in the eastern
nucleus, the alignment of the 150\,MHz features above and below the disk, and
the similarity to the 150\,MHz observations of M\,82, we interpret the 150\,MHz
features as evidence for an outflow in the eastern nucleus, with the southern side being
the closer (approaching) part.

Multiple studies estimate strong magnetic fields strengths of a few mG for the
nuclei in Arp\,220 \citep{lacki+beck2013,yoast-hull2016,torres2004}. In a
medium with density $n_{\text{H}_2}\approx10^{4}$\,cm$^{-3}$ and magnetic
field strength 2\,mG, we expect a cooling time of about 1000\,yrs at GHz
frequencies and slightly less at 150\,MHz \citep[their Fig.~1]{lacki+beck2013}.
If we assume wind speed of 1\,000~km~s$^{-1}$, i.e. higher than estimated by
\cite{sakamoto2009} and \cite{tunnard2015} but less than the 10\,000~km~s$^{-1}$ reached
by SNe \citep{batejat2011}, the CRs could travel only 1\,pc before fading. Even
if the wind speed was as high as 5\,000~km~s$^{-1}$ the CRs could barely escape
the disk (thickness 10\,pc; \citealt{scoville2015}).  The fact that we find
emission as far as 100\,pc ($0.3\farcs$) south of the eastern nucleus means
that the emitting CRs were accelerated tens of parsecs outside the nuclei, due
to star formation outside the nuclei and/or shock-acceleration of CRs in the
outflow.  We note that Arp\,220 has been recently detected in gamma-rays with
Fermi \citep{griffin2016,peng2016}.  The measured $\gamma$-ray spectrum
presented by \cite{peng2016} is roughly an order of magnitude larger than the
predictions of \cite{yoast-hull2015}. This supports the conclusion that a
significant fraction of the radio emission must come from CRs accelerated
outside the nuclei.

It is hard to distinguish between a galactic wind plume driven by star
formation and structure due to an AGN hidden in the eastern nucleus.  Although
the eastern nucleus partially similar to M\,82, which is thought to be
starburst driven, the non-perpendicular direction of the outflow may be a sign
of AGN activity since the M82\, outflow is perpendicular to the disk both in
radio and NIR images.  However, given that Arp\,220 is an ongoing merger, this
discrepancy may be also explained by the interaction forces of the merging
process.  More data is needed to determine what is powering the outflow in the
eastern nucleus.

\subsubsection{An outflow in the western nucleus}

The velocity structure in the CS molecule \citep{scoville2015} suggests a
disk-like rotation for the western nucleus, with a major axis of the disk in
the east-west direction.  If an outflow carries CRs out perpendicular to the
disk, we expect an elongation in the north-south direction, possibly with a
shift of the peak of emission at 150\,MHz  due to the inclination. This is
consistent with the results presented in Sect. \ref{sect:results}, where we see
an elongated feature extending $0.9''$ (330\,pc) south of the western nucleus at
150\,MHz, 1.4\,GHz and 6\,GHz. 

We also find that even though the contribution from extended emission near the
western nucleus is significant, the model still underpredicts the brightness of
this nucleus by 40\% at 150\,MHz, see Fig.~\ref{fig:nuclei}.  This
indicates that there is emission close to this nucleus which is not included in
the model. A part of this emission likely comes from this elongated component.

Although the LOFAR in-band spectral index map (Fig.
\ref{fig:spixinband}) shows a clear north-south elongated feature extending
about $0.9''$ (330\,pc) from the western nucleus, the spectral index maps from
150\,MHz to 1.4\,GHz and 6\,GHz (Figs. \ref{fig:spix150_14} and
\ref{fig:spix150_6}) do not show this feature.
However, if there are free electrons in this region (or in the foreground, for
example in a surrounding star forming disk), free-free absorption could be
important not only for the nuclear disk (as evident from the modelling in Sect. \ref{sect:modeling})
but also for the elongated feature. The turnover frequency where free-free
absorption becomes important depends on the surface brightness (e.g. Eq.
\ref{eqn:condon}). The elongated feature is much weaker than the western nucleus
and consequently would have a lower turnover frequency. In fact, the measured
surface brightness of 40$\times 60\mu$\,Jy~beam$^{-1}=14$\,mJy~arcsec$^{-2}$ at
1.4\,GHz would imply a turnover frequency of about 500\,MHz \citep[their Fig
2]{condon1991}. This could explain the positive spectral index detected within
the LOFAR band, while also being consistent with the flat or negative spectral
index in Figs.  \ref{fig:spix150_14} and \ref{fig:spix150_6}, and consequently
a very low surface brightness at GHz frequencies.

We interpret the elongated feature detected in the western nucleus as evidence of an
outflow, inclined so that the southern part is the approaching
part of the outflow, consistent with the model suggested by \cite{tunnard2015},
their Fig.~16.  We note that our outflow-extension is ten times larger than the
40\,pc reported by \cite{tunnard2015}. This can be explained by
shock-acceleration or star formation also outside the western nucleus, as discussed in Sect.
\ref{sect:eastern} for the eastern nucleus, as we would not expect
CRs accelerated in the western nucleus to keep radiating out to 330\,pc.  We
note that the \emph{HST} NICMOS images of Arp\,220 presented by
\cite{scoville1998} at 1.1\,$\mu$m and 1.6\,$\mu$m also show features north and
south of the western nucleus.  Similarly to the eastern nucleus,
\cite{scoville1998} argue that the southern side is the closer one, based on
the extinction.

\subsection{Extended star formation and shocks in the superwind}
The fact that we see synchrotron emission as far as 1.5\,kpc ($4''$) from the
nuclei in Fig.~\ref{fig:LOFAR} means that the emitting CRs must either
be produced this far from the centre, or they must have been advected from
regions closer to the nuclei by strong winds, possibly driven by the outflows
seen in the nuclei.  A combination of the two effects is also possible.  

There is ample evidence for molecular gas several kpc from the nuclei in the
form of a disk of radius 0.5-1\,kpc, inclination $~45^\circ$ and P.A.
$~45^\circ$ \citep{scoville1997,downes1998,sakamoto1999,koenig2012}.  This
structure match very well the contours of the extended emission in
Fig.~\ref{fig:LOFAR}.  

On arcsecond scales, the extended emission detected at 150\,MHz matches well
the molecular disk, as seen in CO(1-0) Fig.~\ref{fig:co10}, but seem to have
little relation to the orientation of the two nuclei. Although outflows from
the nuclei may contribute to the extended emission, it is likely that a
significant part of it is produced by star formation in the molecular disk
itself.  While the nuclei are very dense within the central 50\,pc
($n_{\mathrm{H}_2}\sim2\times10^5$\,cm$^{-2}$; \citealt{scoville2015}), the
kpc-scale disk is likely less dense and may have weaker magnetic fields
\citep{torres2004}. If so, we expect CRs accelerated here to be able to travel
further than those accelerated in the nuclei, which could explain the smooth
structure of the extended emission. 

\begin{figure*}[htbp]
\centering
\subfigure{
	\includegraphics[width=0.4\textwidth]{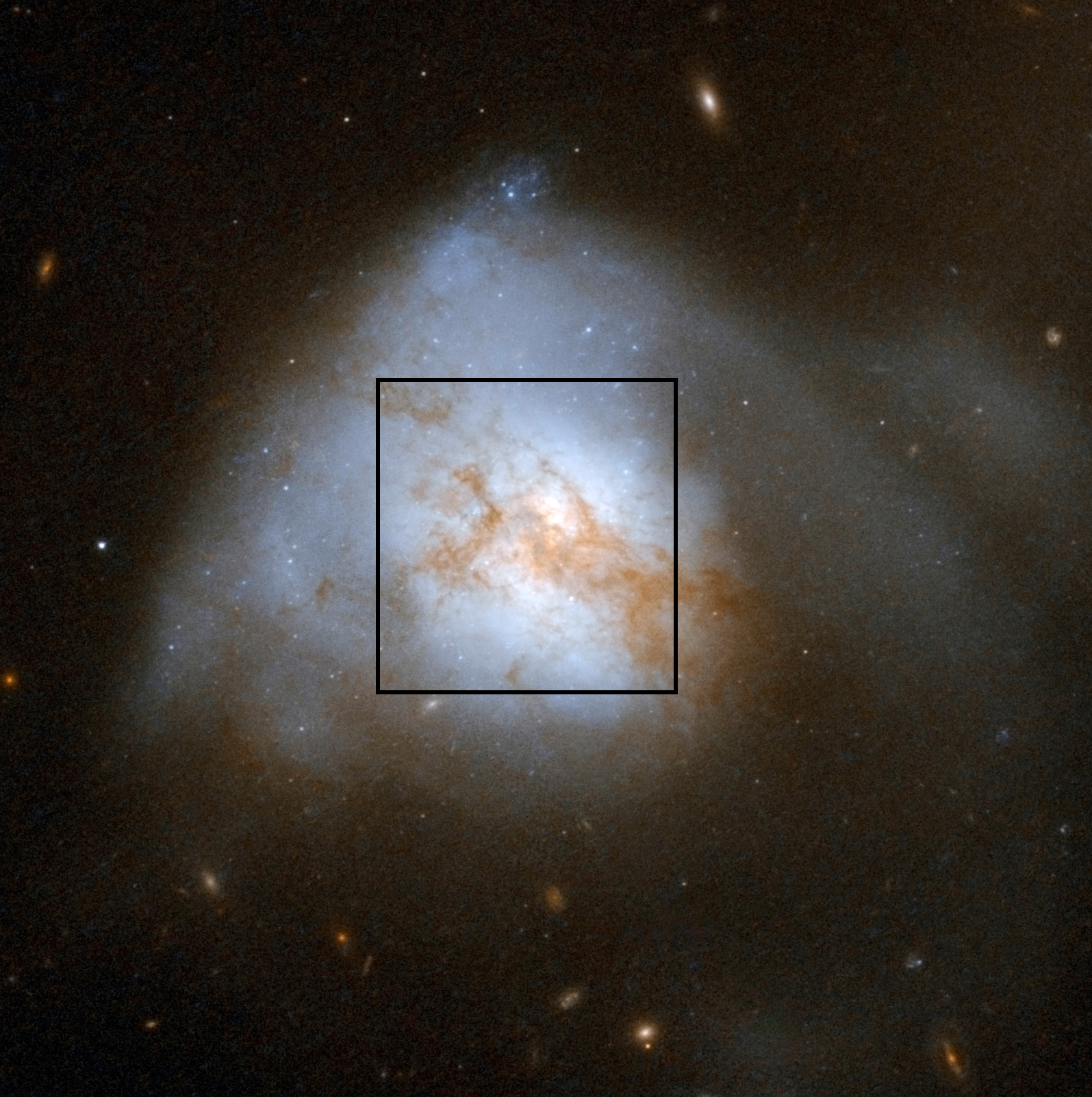}
        \label{fig:HST-RGB}
}
\subfigure{
	\includegraphics[width=0.48\textwidth]{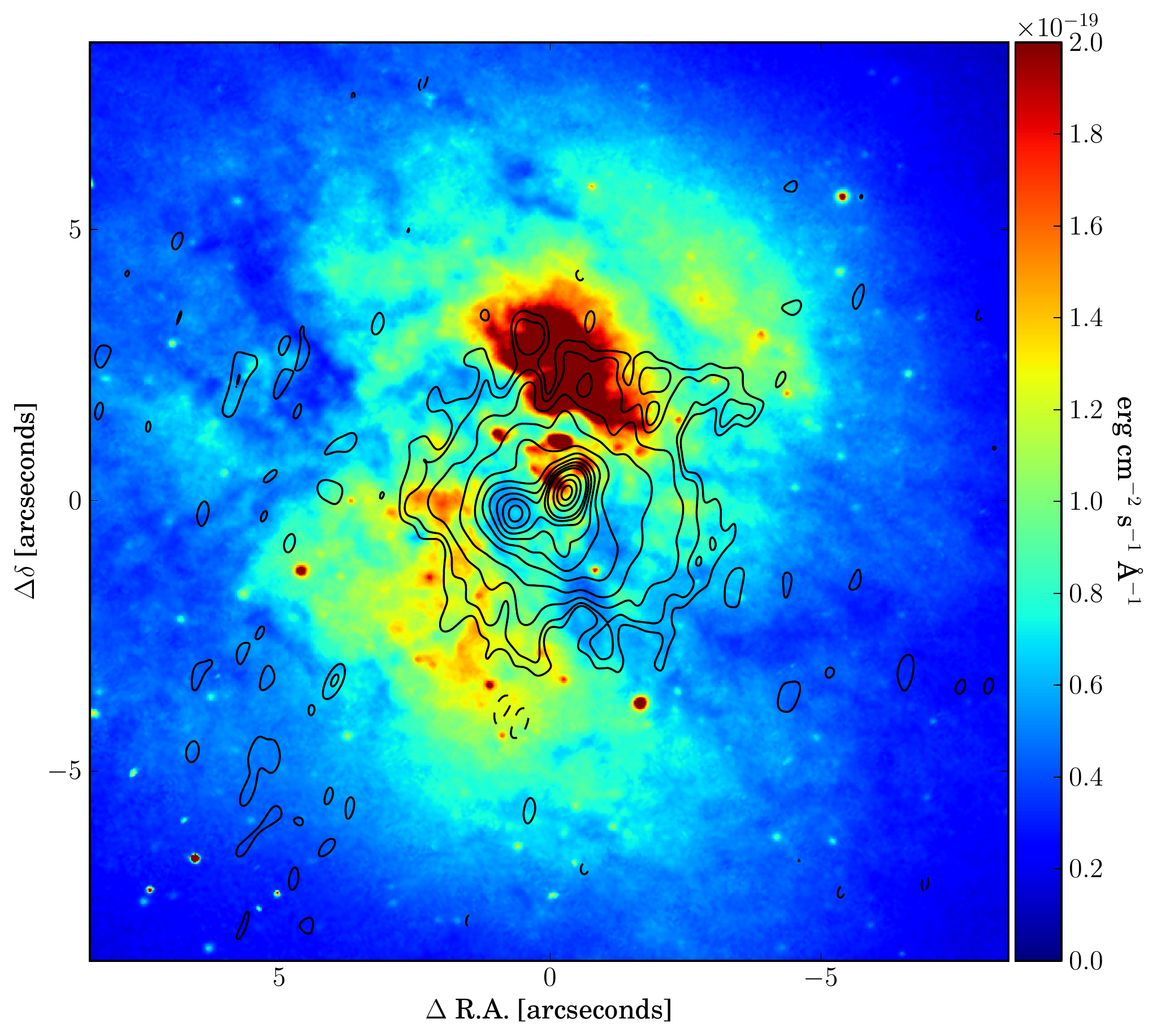}
        \label{fig:HST-I}
}
\caption{The left panel shows the HST ACS colour image previously shown by
\cite{lockhart2015}, where north is up and east is left.  Note the tails on the
west side and the compressed, almost flat, side of the galaxy towards the
north-east, as well as significant extinction by dust towards the centre. The
box has sides of approximately $15''$ and indicates the extent of the right
panel.  The right panel is a zoom showing the F814W (I-band) data with a
saturated colour scale for easy comparison to the results presented in
\cite{arribas2001}, their Fig.~2.  The black contours are the 150\,MHz
continuum at the same levels as in Fig.~\ref{fig:LOFAR}. 
\label{fig:HST}
}
\end{figure*}

Although CRs may be accelerated in the disk, it is hard to see how they could stay 
radiating long enough to reach the ``hooks'' of radio
emission in Fig.~\ref{fig:LOFAR} at 1.5\,kpc from the centre (which would
require a cooling time of $1.5\times10^6$\,yrs assuming a wind speed of
1~000\,km~s$^{-1}$).  However, we note that these ``hooks'' appear perpendicular
to the kpc-scale disk, and are also roughly aligned with the direction of the
outflows from the nuclear disks. We therefore interpret the ``hooks'' as a sign
of a kpc-scale outflow, driven both by the nuclear outflows and by star formation
in the extended disk.
Indeed, Arp\,220 is known to have a large scale bi-conical outflow, or superwind,
seen at optical wavelengths and X-rays \citep{heckman1990,arribas2001,mcdowell2003}. 
For easy comparison with optical data, we show the 150\,MHz emisson as contours
plotted on archival HST I-band data (described in Sect.  \ref{sect:restobs}) in Fig.
\ref{fig:HST}.  
Although the CRs carried from the nuclei may have lost most of their energy at
1.5\,kpc from the centre, the mechanical energy output of the nuclear outflows
may still be the driving force of the superwind. Thus, the wind (and the
magnetic field) would reach much further out than 1.5\,kpc from the centre.
Similarly to the nuclear outflows, shocks in this superwind could accelerate
CRs to produce emission far out from the centre of the galaxy.
We find this a likely explanation of the ``hooks'' in Fig.~\ref{fig:LOFAR}.
We note that observations with even higher sensitivity could potentially detect
even weaker (and more extended) radio emission. Furthermore, observations
with full polarisation calibration could trace the field lines of the outflow
similar to the work by \cite{heesen2011}.

We note that both ``hooks'' appear bent towards the south-west. Such a bending
could be due to movement of the galaxy towards the north-east relative to the
surrounding medium. We note that the HST-image, Fig.~\ref{fig:HST-RGB}, shows a
flattening towards the north-east, consistent with such motion.

\subsubsection{The rate of star formation in the extended disk}
In Fig.~\ref{fig:co10} we show the 150\,MHz radio continuum together with the
CO(1-0) line emission. Let us assume that the $50\sigma$ contours of CO(1-0)
and $5\sigma$ contours at 150\,MHz trace the same extended star forming medium.
Using $N_{\mathrm{H}_2}/W_\text{CO}=1.8$ \citep{dame2001} and Eq. 4 of
\cite{kennicutt1998} we can derive a star formation rate per area of
10\,M$_\odot\text{\,yr}^{-1}$kpc$^{-2}$ from CO(1-0). By extrapolating the
150\,MHz brightness to 1.4\,GHz ($\alpha=-0.7$) and using Eq. 6 of \cite{bell2003} we derive a
value of 2\,M$_\odot\text{\,yr}^{-1}$kpc$^{-2}$.  We note that, in addition to
all uncertainties involved in applying these equations to the dense and complex
environment of Arp\,220, we also expect the CO value to be too high because of
blending with the nucleus, and the 150\,MHz value to be too low because of
synchrotron losses in the outer parts of the disk.  This is consistent with the
difference, and although we refrain from any detailed interpretation of these
numbers, we note that they are consistent within an order of magnitude and
hence do not argue against star formation in the extended disk.


\subsection{Estimating the total star-formation rate}
\label{sect:SFR}
From the flux densities and spectral indices used in the model described in Sect.
\ref{sect:modelingres}, we can extrapolate the radio emission from 32.5\,GHz to
1.4\,GHz without the influence of thermal absorption. We obtain 206\,mJy (east),
281\,mJy (west) and 73.2\,mJy (sphere), i.e. a total spectral luminosity of
$L_{1.4\text{\,GHz}}=3.99\times10^{23}\text{W~Hz}^{-1}$.  Using Eq. 6 by
\cite{bell2003} we calculate the star formation rate (SFR) as
$5.52\times10^{-22}L_{1.4\text{\,GHz}}\approx220$\,M$_\odot\text{\,yr}^{-1},$
in good agreement with the 240$\pm30$\,M$_\odot\text{\,yr}^{-1}$ calculated by
\cite{farrah2003} using the far-infrared luminosity. Scaling the SFR by the
flux density of each component at 1.4\,GHz, we find the three components contributing
81\,M$_\odot\text{\,yr}^{-1}$ (east), 110\,M$_\odot\text{\,yr}^{-1}$ (west) and
29\,M$_\odot\text{\,yr}^{-1}$ (sphere).

\subsection{Arp\,220 follows the FIR/radio correlation}
\label{sect:FIRradio}
The mean ratio of far-IR (FIR) emission to radio emission is usually quantified
as a logarithmic ratio called the $q$ parameter (e.g. \citealt{yun2001}, their Eq. 5).
\cite{yun2001} find Arp\,220 to have $q=2.67$ which is larger
than the mean of $2.34$ for the IRAS 2~Jy sample containing over 9000
sources, i.e.  showing less radio emission than expected for Arp\,220. 

However, using the 1.4\,GHz flux density extrapolated in Sect. \ref{sect:SFR}
together with IR measurements from the IRAS point source catalogue
v2.1\footnote{Obtained via http://irsa.ipac.caltech.edu/Missions/iras.html.} of
$S_{60\mu\text{m}}$ = 104.1\,Jy and $S_{100\mu\text{m}}=117.7$\,Jy we obtain
(using Eqns. 5 and 6 by \citealt{yun2001}) a value of $q=2.36$ for Arp\,220,
closer to the average value.  This shows the importance of accounting for
thermal absorption even at GHz frequencies. The fact that Arp\,220 follows the
FIR/radio correlation indicates that it is powered by star formation rather
than AGN activity. This is consistent with the large number of compact
supernovae and supernova remnants found with VLBI observations (see e.g.
\citealt{batejat2011}).

In addition to star formation and AGN activity, tidal shocks in
merging systems may heat the dust and gas enough to produce additional FIR and
synchrotron emission, thereby possibly affecting the FIR/radio correlation
\citep{murphy2013,donevski2015}.  \cite{donevski2015} quantify this effect in
terms of the $q$-parameter for different classes of mergers, using the same
six-stage merger classification scheme as \cite{haan2011} based on HST imaging.
We note that Arp\,220 is hard to classify based on optical imaging due to the
extreme dust obscuration of the centre. \cite{haan2011} and \cite{murphy2013} classify
Arp\,220 differently using the same scheme, as class 6 and 5 respectively. Both
these classes are describing a post-merger system with a single nucleus.  We
know, from radio observations that Arp\,220 has two nuclei, although the
separation of 370\,pc is much smaller that the median ULIRG separation of
1.2\,kpc found by \cite{haan2011} and is therefore easily missed in the HST
classification scheme.  If we re-classify Arp\,220 as class 4, a late ongoing
merger with double nuclei and a tidal tail, our measured $q$-value is in good
agreement with the $2.31\pm0.17$ expected for this class by
\cite{donevski2015}. Hence, by this argument, emission from tidal shocks seem
to play a minor role in Arp\,220.

\section{Summary and outlook}
\label{sect:summary}
Using the international LOFAR telescope we obtain, for the first time, an image
of Arp\,220 at 150\,MHz with subarcsecond resolution.  We detect emission
associated with the two nuclei known from GHz frequencies, but also extended
radio emission reaching 1.5\,kpc from the nuclei.  The nuclei have positive
spectral indices, indicating significant free-free absorption, while the
extended emission has an overall spectral index of $-0.7$, typical for
optically thin synchrotron radiation.  We find that the extended emission
accounts for more than 80\% of the total flux density measured at 150\,MHz.

We report on elongated features in the two nuclei, extending 0.3'' (110\,pc)
from the eastern nuclear disk and $0.9''$ (330\,pc) from the western nuclear
disk, and we interpret these features as evidence for outflows.  

The extended radio emission follows the CO(1-0) distribution and is likely
coming from star formation in the kpc-scale molecular disk surrounding the two
nuclei.  Outflows from this disk, as well as from the nuclei, likely drives the
superwind seen in optical and X-rays wavelengths. 

We find that shock-acceleration of CRs in the outflows, both in the nuclei and
in the base of the superwind) is required to explain the extent of the radio
emission.

We model the Arp\,220 as a three-component model: the nuclei as exponential
disks with thermal absorption, surrounded by a uniform sphere of optically thin
synchrotron emission. Our model successfully explains the basic shape of the
observed integrated spectrum of the galaxy, as well as the spectra of the
nuclei, which are well described by a mixed thermal/non-thermal plasma of
thermal fractions 0.8\% and 0.4\% at 1 GHz for the east and west nucleus
respectively. These values are an order of magnitude lower than the $9\%\pm3\%$
found by \cite{marvil2015} for a large sample of galaxies. Still, our thermal
fractions for Arp\,220 are upper limits because we assume no foreground thermal
absorption.  The low thermal
fractions may be explained by dust absorbing a major part of the ionising
photons produced by young stars. 

Our model underpredicts the flux density in the range 200\,MHz to 1\,GHz for
Arp\,220, even when including extended emission, indicating that the emission
from outflows in the nuclei (which are prominent below 1\,GHz, but are not
considered in our simple model) play an important role in this galaxy.

When including the extended emission and accounting for absorption effects, we find
that Arp\,220 follows the FIR/radio correlation with $q=2.36$, and we estimate
a total star formation rate of 220~M$_\odot\text{\,yr}^{-1}$.

International LOFAR observations show great promise to detect extended
structures such as outflows or radio halos at MHz frequencies, where they are
bright due to the synchrotron spectral slope and lack of free-free absorption.
Future international LOFAR observations of Arp\,220 using the Low Band Array at
60\,MHz would be very useful to further disentangle the contributions of the
different emitting structures in Arp\,220.

Dutch-LOFAR observations will in the future be used to detect and study star
forming galaxies. Our results show that for LIRGs the emission detected at
150\,MHz does not necessarily come from the main regions of star formation.
This implies that unresolved observations of such galaxies at 150\,MHz is
unlikely to be useful for deriving star formation rates. Future studies of
LIRGs at MHz frequencies would therefore benefit from using international LOFAR
baselines to resolve the star forming structure.

\begin{acknowledgements}
E.V. acknowledges support from the Royal Swedish Academy of Sciences.
A.A. and M.A.P.T. acknowledge support from the Spanish MINECO through
grants AYA2012-38491-C02-02 and AYA2015-63939-C2-1-P, partially funded by
FEDER funds.
LOFAR, the Low Frequency Array designed and constructed by ASTRON, has
facilities in several countries, that are owned by various parties (each with
their own funding sources), and that are collectively operated by the
International LOFAR Telescope (ILT) foundation under a joint scientific
policy. We note the valuable assistance provided by the LOFAR Science Support
during this work.
The research leading to these results has received funding from the European
Commission Seventh Framework Programme (FP/2007-2013) under grant agreement No
283393 (RadioNet3).
e-MERLIN is the UK's National Radio Interferometric facility, operated by the
University of Manchester on behalf of the Science and Technology Facilities
Council (STFC).
Using data from the Karl G. Jansky Very Large Array (VLA), we acknowledge
that the National Radio Astronomy Observatory is a facility of the National
Science Foundation operated under cooperative agreement by Associated
Universities, Inc.
	This research has made use of the NASA/IPAC Extragalactic Database (NED)
which is operated by the Jet Propulsion Laboratory, California Institute of
Technology, under contract with the National Aeronautics and Space
Administration.
One image presented in this paper was obtained from the Mikulski
Archive for Space Telescopes (MAST). STScI is operated by the Association of
Universities for Research in Astronomy, Inc., under NASA contract NAS5-26555.
Support for MAST for non-HST data is provided by the NASA Office of Space
Science via grant NNX09AF08G and by other grants and contracts.
\end{acknowledgements}

\bibliographystyle{aa}
\bibliography{allrefs}

\appendix
\section{Calibration of LOFAR data}
\label{app:lofarcal}
The observations, calibration and imaging of the LOFAR data were carried out
based on experience from earlier work on M82 \citep{varenius2015}.  These data
on Arp\,220 were observed in project LC2\_042 (P.I.: E. Varenius) between
18:30 and 02:15 UT on June 4th 2014 with six hours on Arp\,220.  The observations
included 46 LOFAR high band array (HBA) stations: 24 core stations (CS) in
\emph{joined} mode (where the two 24-tile ``ears'' of the station are added to
form a single station), 14 remote stations (RS), and eight international
stations (IS).

The total available bandwidth was split equally in two simultaneous beams of
width 48\,MHz (240 sub-bands), centred on 150\,MHz; one on Arp\,220 and one on
the bright and compact calibrator J1513+2388 separated from Arp\,220 by
$4.9^\circ$.  Every 20 minutes the observations switched, for 2 minutes, to a
single beam on the absolute flux density calibrator 3C295, separated from
Arp\,220 by 32$^\circ$.  
	Although in theory a single scan of 3C295 is sufficient to determine the
	flux scale, we included scans every 20 minutes for two main reasons.
	Firstly: in case the primary delay calibrator would have been very weak, we
	could have used the scans of 3C295 to track the relative phase of the CS to
	coherently combine the CS visibilities into a larger super-station with
	enough sensitivity to international stations to derive delay solutions
	towards a weak calibrator. However, J1513+2388 was strong enough to derive
	delay and rate solutions without phasing up the core. Secondly: the current
	LOFAR beam model is not accurate enough to correctly transfer the flux
	scale (and spectral index) over large (typically more than 10 degrees)
	elevation differences. Tracking both J1513+2388 (close to the target) and
	3C295 (further away) through a wide range of relative elevation angles made
	it possible to measure the gain differences as a function of elevation
	separation during the observation. This enables us to choose a good
	time range for flux calibration, i.e. when 3C295 is close in elevation to
	J1513+2338, as well as estimate the flux error due to residual beam
	effects. After careful investigations, we are confident that our flux scale
	and LOFAR in-band spectral index are correct within the uncertainties
given in this paper.

The correlated data were stored in the LOFAR long term
archive with visibility sample resolution 2\,seconds in time and 48.8\,kHz in
frequency (i.e. 4 channels per LOFAR HBA sub-band).  This resolution defines a
field of view due to coherence loss on long baselines (smearing).  We estimate the
coherence loss due to smearing for a 1000\,km baseline to be at most 45\% for
sources at $10'$ distance from the phase centre. 

All measurement sets were processed with the LOFAR \emph{ default
pre-processing pipeline} (DPPP) to edit bad data using AOflagger
\citep{offringa2012}, and were corrected for the array and element beam response.
For processing speed, the data were then further averaged to a sample
resolution of 10\,seconds and 195\,kHz. For clarity we note that we did not use
the \emph{mscorpol} model as done by \cite{varenius2015} to correct for the
LOFAR beam since mscorpol does not include a frequency dependence.  For the
wide bandwidth in this observation, we considered the LOFAR beam model more
appropriate, although preliminary investigations show these two models to have
similar accuracy.  The data were then converted to circular polarisation using
the table query language (TAQL), and to UVFITS format using CASA. 

Similar to \cite{varenius2015}, residual delays and rates were corrected for
all non-core stations by using baselines longer than 60~k$\lambda$ for
J1513+2338, where the total bandwidth was divided in eight groups (AIPS IFs) of
5.9\,MHz each for a piece-wise linear approximation of the dispersive delay to
be valid. The task \verb!FRING! was used with a solution interval of 2 minutes. 
Typical delay values were tens of ns for RS and 100-300\,ns for IS,
with typical differences of about 15\% between the lower and upper IFs.

The corrections were applied to all sources, thereby removing major
residual errors due to the ionosphere. The data could now be further averaged
in time and frequency, and for processing speed we averaged to a sample
resolution of 30 seconds and 1.2\,MHz.  After averaging, the visibility weights
were re-calculated to reflect the scatter within 5 minute solution intervals.  

\subsection{Excluded stations}
The Swedish LOFAR station in Onsala, SE607, and the core station CS501 were excluded 
from calibration and imaging.  Losing CS501 has only a very
minor impact on sensitivity. However, losing SE607 means significantly lower
resolution in north-south direction in the final synthesised image.  Although
derived delay and rate corrections towards J1513+2338 looked reasonable for
SE607, we found a systematic reduction in visibility amplitudes on parts
of the bandwidth to this station, where the upper and lower halves of the
bandwidth had significant gain differences. Despite significant efforts to
understand this discrepancy, we found obvious amplitude errors in the final
image when including SE607. We therefore decided to exclude this station,
despite the loss in resolution. More advanced calibration methods may be able
to include the data from SE607, hence gaining resolution in north-south
direction.  This would be valuable to investigate the outflows in the nuclei
of Arp\,220.

\subsection{Flux calibration}
\label{sect:fluxcal}
Amplitude corrections were derived for each antenna, one correction per
spectral window every minute, assuming J1513+2338 to be a point source of
1.55\,Jy at the lower end of the band, with spectral index $\alpha=+0.8$,
consistent with total flux density measurements available via
NED\footnote{http://ned.ipac.caltech.edu/}. To reduce the effect of interfering
sources in the field, only baselines longer than 60~k$\lambda$ were used to
derive amplitude corrections. Multiple rounds of self-calibration were used to
take into account any structure present in the amplitude calibrator, although
we found this source to be point-like at this resolution. The amplitude
corrections were median window filtered and smoothed to correct obvious
outliers using a filter width of 2 hours, before applying the corrections to
all sources.  

A band pass calibration was done to correct for curvature within each
6\,MHz spectral window using J1513+2338 with $\alpha=+0.8$, baselines
longer than 60~k$\lambda$, and a solution interval of 60 minutes.

The absolute amplitude scale, and the in-band spectral index, was checked by
transferring all corrections to 3C295 and, after a phase-self calibration
assuming a point source model for the shortest baselines 0.1-15\,k$\lambda$,
one image was produced for each AIPS IF using these baselines.  The recovered
CLEAN model flux density for each IF was found to be within a few percent of
the model by \cite{scaife2012}.

For imaging of 3C295 we used data from the last 3.5 hours of observing, when
3C295 and Arp\,220 are separated by 10$^\circ$ in elevation.  Based on the flux
density variations of J1513+2338 during the observation we conclude that
inaccuracies in the current LOFAR beam model may introduce systematic flux
density errors of 10\% between sources separated by 10$^\circ$ in elevation.
We therefore adopt a flux density uncertainty of 15\% to account for beam model
errors as well as errors when measuring the flux of 3C295.  

\subsection{Phase calibration}
\label{sect:selfcal}
Because of the large ($4.9^\circ$) separation between J1513+2388 and Arp\,220,
the residual phase errors differ substantially. After phase referencing to
J1513+2388 no peak could be found in the Arp\,220 image, and hence phase
self-calibration was necessary. Self-calibration without a prior knowledge of the
peak position will align the data arbitrarily in R.A. and Dec., and therefore the
image has to be aligned by comparing positions of some alignment sources, as described
later in Sect. \ref{sect:posacc}.

Arp\,220 was phase self-calibrated (in AIPS) using a point source as starting model.
We found Arp\,220 to be too weak for calibration using only international
baselines, mainly due to the limited sensitivity of the CS.
We decided to include shorter baselines, longer than 4~k$\lambda$, in the
self-calibration process, i.e. including also CS-RS (but no CS-CS) baselines.
A solution interval of 1 minute was used and corrections were found for all
AIPS IFs separately to correct residual phase errors between the IFs, although
the image used in each self-calibration step was made from all IFs together. We
note that the phase corrections derived for all IFs were very
similar.

We performed multiple rounds of hybrid mapping until the process converged to a
stable result. We note, however, that self-calibration including short LOFAR
baselines may cause the calibration process to violate the constraints defined
by the closure phases on baselines between international stations, since these
stations have no short baselines and the NL-stations are all relatively close.
We plotted the final CLEAN model against the self-calibrated data and found
that the model was in good agreement with the data also on baselines between
international stations. This means the self-calibration procedure is
constrained by closure phases on all baselines and the results are robust.

We also made the additional test of starting the process of hybrid mapping
using a MERLIN-only 1.4\,GHz image as initial model, i.e. a double source
instead of a point.  From this we also recovered the same final structure, i.e.
the final result does not depend on the initial model used for
self-calibration.

The derived phase corrections were applied to Arp\,220 and to the three
alignment sources listed in Tab. \ref{tab:targetlist}. With residual
phase-variations corrected, the data could be further averaged to a sample
resolution of 120 seconds in time for processing speed.  Given this, we
estimate our field of view around Arp\,220 to be limited to within a radius of
10$''$ (10\% coherence loss on a 1000\,km baseline).

\subsection{Imaging and sensitivity}
An image of Arp\, 220 were obtained using the multi-scale (MS)
multi-frequency-synthesis (MFS) CLEAN algorithm as implemented in CASA using 2
Taylor terms, and baselines longer than ~2\,k$\lambda$.  The scales used were: a
point, and two Gaussians of FWHM 1$''$ and 4$''$. The final RMS noise level
was 0.15\,mJy~beam$^{-1}$ with a CLEAN restoring beam of $0\farcs65\times0\farcs35$. We
also obtained an in-band spectral index map as constructed by the
MFS-algorithm.  

To check the thermal noise we obtained an image of Arp\,220 in Stokes V. This
image contained only noise with RMS 0.14\,mJy~beam$^{-1}$, suggesting that
the stokes I image, with very similar sensitivity, is limited random noise and
not by residual phase errors or interfering sources in the field.  The achieved
sensitivity can be compared to the 0.15\,mJy~beam$^{-1}$ achieved by
\cite{varenius2015} for M82 at 154\,MHz, using 16 hours of data with 16\,MHz
bandwidth. Scaling the M82 noise to 6 hours and 48\,MHz bandwidth, we expect
0.14\,mJy~beam$^{-1}$, in excellent agreement with the observed value.

\subsection{A plan for alignment of the LOFAR data}
To maximise the sensitivity on Arp\,220, the available bandwidth was split between
only two beams, i.e. no separate beam on a nearby phase calibrator as
done by \cite{varenius2015}.  Instead, we aimed to align the final LOFAR image
with observations with similar resolution at higher frequencies, e.g. MERLIN at
1.4\,GHz. However, the morphology seen with LOFAR was different from MERLIN
(especially the eastern nucleus), making manual alignment using Arp\,220 itself
very challenging. 

\begin{figure*}
\centering
\includegraphics[width=\textwidth]{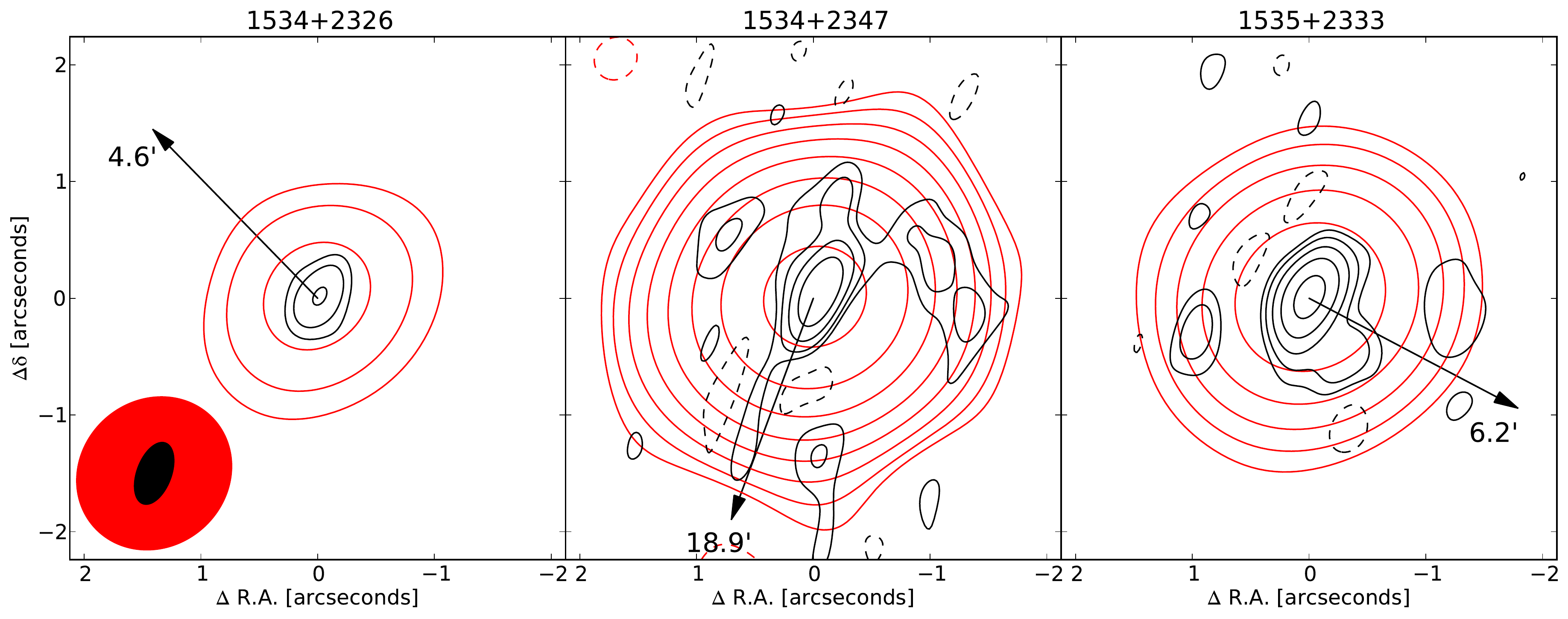}
\caption{The three in-beam sources detected both at 1.4\,GHz and 150\,MHz, after
applying the shift derived from 1534+2326 (left panel) and 1534+2333 (right panel)
to all sources. Red contours show VLA 1.4\,GHz as [-5,5,10,20,40,80,160,320,640]$\times$0.05\,mJy~beam$^{-1}$, 
and black contours show LOFAR 150\,MHz as [-5,5,10,20,40,80,160]$\times$0.15\,mJy~beam$^{-1}$. The black
arrow shows the direction towards Arp\,220 with the respective angular distance
at the tip.  The middle source is clearly affected by smearing at 150\,MHz,
and was therefore not included in the alignment as described in the text.
the CLEAN restoring beams of VLA (red: $1\farcs38\times1\farcs24$) and LOFAR (black:
$0\farcs65\times0\farcs35$) are shown in the lower left. }
\label{fig:cals}
\end{figure*}

However, we detect multiple background sources within the smearing radius
defined by the time and frequency resolution of the LOFAR data. Unfortunately
there is not yet a catalogue available with positions accurate enough at LOFAR
frequencies to align the data.  Instead, we sought compact background
sources in the field which were seen both in the archival 1.4\,GHz VLA data,
described in Sect. \ref{sect:vladata}, and in the LOFAR data. Although single
sources may have frequency-dependent positions, the use of multiple sources
provides a robust way of aligning the LOFAR data to the astrometric reference
frame of the VLA observations. 

We note that in the future it may be easier to find close, bright and compact
calibrators for phase referencing at 150\,MHz by using the upcoming LOBOS
catalogue \citep{lobos2015}.

   \begin{table}
      \caption[]{Positions for alignment sources}
         \label{tab:alignment}
         \begin{tabular}{ l l l}
            Source      &  R. A. [J2000] & Dec. [J2000]\\
            \hline
            1534+2326 \tablefootmark{a}& $15^{\rm h}34^{\rm m}43^{\rm s}.2670$ & $23^\circ26'54\farcs146$\\
            1534+2347 \tablefootmark{a}& $15^{\rm h}34^{\rm m}28^{\rm s}.6793$ & $23^\circ47'52\farcs427$\\
			1535+2333 \tablefootmark{a}& $15^{\rm h}35^{\rm m}21^{\rm s}.3291$ & $23^\circ33'05\farcs664$\\
            \noalign{\smallskip}
            \hline
         \end{tabular}
         \tablefoot{
			 \tablefoottext{a}{Position obtained from archival VLA data at
			 1.4\,GHz, see Sect. \ref{sect:posacc} for uncertainties.}
}
   \end{table}

\subsection{Finding alignment sources in the VLA data}
To find sources suitable for alignment of the
150\,MHz image, we first imaged the VLA data described in Sect.
\ref{sect:vladata} using the multi-scale CLEAN algorithm as
implemented in CASA 4.5.2, and obtained a model image of Arp\,220 with resolution
$1\farcs38\times1\farcs24$  and RMS noise 50\,$\mu$Jy beam$^{-1}$.  
Given the time and frequency resolution of these data we estimate coherence
losses of 20\% on the longest (36\,km) VLA baselines at a distance of $10'$
from the phase centre. Smearing is hence a greater limitation than the primary
beam of about $30'$ for a VLA antenna at 1.4\,GHz, when imaging sources far
away from the phase centre.  

The CLEAN model for Arp\,220 was then UV-subtracted from the data to enable
cleaning of faint nearby sources and a big image was made to identify compact
sources. Three sources were selected as potential alignment sources and images
were obtained for each of them. By fitting Gaussian intensity distributions to
the alignment sources we obtained the positions and flux densities at 1.4\,GHz, 
see Tables \ref{tab:alignment} and \ref{tab:alignmentflux}. The fitting
uncertainties reported by CASA were 15\,mas, 2\,mas and 3\,mas respectively for
the sources 1534+2326, 1534+2347 and 1535+2333 at 1.4\,GHz.

\subsection{Aligning the LOFAR data}
For minimal time and bandwidth smearing of the three alignment sources in the LOFAR
data, we phase-shifted the Arp\,220 data with resolution (2s and 48.8\,kHz) to the
positions of the three sources before applying the same cumulative phase and
amplitude corrections as applied to Arp\,220, i.e. including the corrections
derived during the self-calibration procedure described in Sect.
\ref{sect:selfcal}.  
Unfortunately, the brightest source, 1534+2347, we found to be too severely
affected by smearing to be used for positional alignment, but we include
position (Table \ref{tab:alignment}) and flux density (\ref{tab:alignmentflux})
of this source for future reference as this may be a good reference for future
observations of Arp\,220.

   \begin{table}
      \caption[]{Integrated flux densities for the alignment sources.}
         \label{tab:alignmentflux}
         \begin{tabular}{ l l r l r}
			 Source &  Obs. freq& Resolution & RMS & Int. flux\\
						  &  [GHz] & [arcsec] & [mJy/b]& [mJy]\\
            \hline
			1534+2326 & 0.15 & $0.65\times0.35$ &0.14 & 3.7$\pm0.2$\\
			1534+2326 & 1.4  & $1.38\times1.24$ &0.04 & 1.4$\pm0.1$\\
			1534+2347 \tablefootmark{a}& 0.15 & $0.65\times0.35$ &0.18 & 13.7$\pm0.7$\\
			1534+2347 & 1.4  & $1.38\times1.24$ &0.06 & 43.6$\pm0.2$\\
			1535+2333 & 0.15 & $0.65\times0.35$ &0.17 & 20.7$\pm1.4$\\
			1534+2333 & 1.4  & $1.38\times1.24$ &0.05 & 7.7$\pm0.1$\\
            \noalign{\smallskip}
            \hline
         \end{tabular}
		 \tablefoot{ Flux densities from Gaussian fitting.
\tablefoottext{a}{Possibly reduced due to smearing}. For positions, see Table
\ref{tab:targetlist}.}
   \end{table}

The remaining two sources, 1534+2326 and 1535+2333 were fitted
with a two dimensional Gaussian intensity distribution to obtain positions at
150\,MHz, and CASA reported fitting uncertainties of 13\,mas and 6\,mas respectively.
A common shift was derived to minimise the difference of both positions
relative to the VLA positions, and after shifting the difference between the positions
were less than 5\,mas in R.A. and 2\,mas in Dec. between 1.4\,GHz and 150\,MHz.

\subsection{Positional accuracy}
\label{sect:posacc}
The absolute positional accuracy of our LOFAR images depends both on the relative accuracy
of the alignment to 1.4\,GHz, and on the absolute accuracy of the 1.4\,GHz positions. 

As a conservative estimate on the relative accuracy, we add the error margins
involved in the alignment described above for one single source, although
having two sources should make this error smaller.  The fitting accuracy is
$<15$\,mas at 1.4\,GHz and $<13$\,mas at 150\,MHz. Adding the final alignment
difference of $<5$\,mas gives a relative uncertainty of $<33$\,mas in each
coordinate.

The uncertainty of the absolute reference position of the 1.4\,GHz data depends
on the uncertainty of the phase-referencing to the calibrator J1513+2338.
Inspection of the phase corrections derived for J1513+2338 during calibration
of the 1.4\,GHz VLA data suggests the phases are slowly varying and we estimate errors
in the temporal interpolation between target and calibrator to be negligible.
Although the VLBI-position in Table \ref{tab:targetlist} in theory may be
accurate to less than 1\,mas, there may still be larger sources of uncertainty
due to the spatial extrapolation of the phase solutions during the VLA
observations, as well as effects of structure in the calibrator source. 

From inspection of the phase solutions, the phase on J1513+2338 changes on
average 0.4$^\circ$ per degree elevation. For a median VLA A-array baseline of
12\,km we estimate a positional uncertainty due to linear spatial extrapolation
over a target/calibrator separation of 4.5$^\circ$ of 20\,mas. We note that the
calibrator and target are above 60 degrees elevation for most of the
experiment, a region where the tropospheric mapping function ($\approx$ cosec(el))
is essentially flat, and hence a linear extrapolation is a conservative
estimate of the uncertainty due to spatial extrapolation.

From the FITS images available via http://astrogeo.org/calib/search/html, we find 
that although J1513+2338 is core-dominated at the available frequencies of 2.3\,GHz and 
4.8\,GHz, it has a jet extending 60\,mas south and 30\,mas west from the core. Because of the
core-shift effect, the calibrator position at 1.4 GHz\,may be shifted along the
jet direction with respect to the catalogue position from higher frequencies. To
estimate the magnitude of this shift we measure the relative flux densities of the core and the
jet at 2.3\,GHz and 4.8\,GHz from the FITS images. The core
flux is measured by fitting a Gaussian intensity distribution. The jet flux is
measured by summing the pixels brighter than 3$\sigma$ and subtracting the core
flux. We find that the jet contributes 15\% of the total flux density at 4.8\,GHz
and 16\% at 2.3\,GHz. However, this is likely a lower limit because the jet may
have significant flux density in extended emission with surface brightness
below 3$\sigma$. Furthermore, the core may suffer significant free-free absorption
at 1.4\,GHz, while the extended emission would have a steeper spectrum and hence be 
more important at 1.4\,GHz. Indeed, spectral index of this source is found to be positive
at 150\,MHz, see Sect. \ref{sect:fluxcal}, consistent with significant absorption
of the core. Finally, the core and jet may be time variable.
To account for these effects, we assume an upper limit of the core/jet flux
density ratio of 50\% at 1.4\,GHz. If all the jet flux is at the tip of the jet,
this implies a conservative positional uncertainty of 15\,mas in R.A. and 30\,mas in Dec.

Adding all the above uncertainties, we estimate our 150\,MHz positions for Arp\,220 to
be accurate within 68\,mas in R.A. and 83\,mas in Dec.

\end{document}